\documentclass[english,11pt,aps,prd,a4paper,preprintnumbers,floatfix,nofootinbib,showpacs,superscriptaddress, notitlepage]{revtex4-1} 
 \pdfoutput=1
\usepackage{color}
\usepackage{graphicx}
\usepackage{caption}
\usepackage{subcaption}
\usepackage{amsmath}
\usepackage{amssymb}
\usepackage[colorlinks=true,citecolor=darkred,urlcolor=darkred, pdfborder={0 0 0}]{hyperref}
\usepackage[normalem]{ulem}

\definecolor{darkred}{rgb}{0.6,0,0}

\definecolor{linkcolor}{rgb}{0,0,0.5}

\def\gsim{\raise0.3ex\hbox{$\;>$\kern-0.75em\raise-1.1ex\hbox{$\sim\;$}}}
\def\lsim{\raise0.3ex\hbox{$\;<$\kern-0.75em\raise-1.1ex\hbox{$\sim\;$}}}

\def\beqn#1{\begin{equation}\label{#1}}
\def\eeqn{\end{equation}}

\def\beqa#1{\begin{eqnarray}\label{#1}}
\def\eeqa{\end{eqnarray}}

\def\Z2{$\mathcal{Z_2}$}

\newcommand {\ignore}[1]{}

\newcommand{\sm}{{Standard Model }}

\newcommand{\AddrAHEP}{%
  AHEP Group, Institut de F\'{i}sica Corpuscular --
  CSIC/Universitat de Val\`{e}ncia, Parc Cient\'ific de Paterna.\\
 C/ Catedr\'atico Jos\'e Beltr\'an, 2 E-46980 Paterna (Valencia) - SPAIN}

\begin{document}

\bibliographystyle{unsrt}   

\title{\boldmath Testing generalized CP symmetries with precision studies at DUNE}
\author{Newton Nath}
\email[Email Address: ]{newton@ihep.ac.cn}
\affiliation{
Institute of High Energy Physics, Chinese Academy of Sciences, Beijing, 100049, China}
\affiliation{
School of Physical Sciences, University of Chinese Academy of Sciences, Beijing, 100049, China}
\author{Rahul Srivastava}
\email[Email Address: ]{rahulsri@ific.uv.es}
\affiliation{\AddrAHEP}
\author{Jos\'{e} W. F. Valle}
\email[Email Address: ]{valle@ific.uv.es}
\affiliation{\AddrAHEP}

\begin{abstract}
  \vspace{1cm}

We examine the capabilities of the DUNE experiment in probing leptonic CP violation within the framework of theories with generalized CP symmetries characterized by the texture zeros of the corresponding CP transformation matrices. We investigate DUNE's potential to probe the two least known oscillation parameters, the atmospheric mixing angle $\theta_{23}$ and the Dirac CP-phase $\delta_{\rm CP}$. We fix theory-motivated benchmarks for ($ \sin^2\theta_{23}, \delta_{\rm CP} $) and take them as true values in our simulations. Assuming 3.5 years of neutrino running plus 3.5 years in the antineutrino mode, we show that in all cases DUNE can significantly constrain and in certain cases rule out the generalized CP texture zero patterns.

\end{abstract}

\maketitle


\section{Introduction}


The discovery of oscillations~\cite{Kajita:2016cak,McDonald:2016ixn} provides a major milestone in the development of 
particle physics over the past few decades~\cite{Valle:2015pba}. 
Several fundamental open issues in cosmology may also be closely related to the lepton sector and the properties of neutrinos, which we are just starting to uncover. 
For example, neutrinos could hold the key to the mystery associated to the origin of the baryon asymmetry of the universe.
Although this could arise within the \sm through anomalous electroweak baryon-number non-conserving processes~\cite{Kuzmin:1985mm}, the mechanism on its own turns out not to be realistic~\cite{Dine:2003ax}. 
However, sphaleron processes can convert a pre-existing lepton number, producing a net baryon number, a process called leptogenesis~\cite{Fukugita:1986hr}. 
Indeed, this mechanism could in principle account for the observed matter to anti-matter asymmetry of the universe provided CP is violated in the lepton sector in an adequate manner.
This brings the issue of CP violation in the neutrino sector to the spotlight. 
By exploring the phenomenon of neutrino oscillations, the Deep Underground Neutrino Experiment (DUNE)~\cite{Acciarri:2016crz} aims to improve our understanding of neutrinos through the study of one of the three CP phases present in the simplest theories of massive neutrinos~\cite{Schechter:1980gr}.

DUNE is an international experiment for neutrino studies that will consist of two neutrino detectors placed in the world's most intense neutrino beam. 
The Long-Baseline Neutrino Facility will provide the neutrino beamline, while the two detectors will play complementary roles. 
These detectors will have the capability of searching for new CP violating features of neutrino oscillations, thereby probing the existence of leptonic CP violation.
There have already been symmetry-based studies in the context of the DUNE experiment performed, for example, in Refs.~\cite{Chatterjee:2017ilf,Srivastava:2017sno,Srivastava:2018ser,Chakraborty:2018dew,Nath:2018xkz}.
In particular, we note that popular flavor symmetry frameworks with $\mu - \tau$ symmetry, such as those in~\cite{Fukuyama:1997ky,Babu:2002dz,Harrison:2002et,Grimus:2003yn} (see Ref.~\cite{Xing:2015fdg} for a recent review and references) usually predict $\theta_{13}=0$ (zero reactor mixing), and also $\theta_{23}, \delta_{\rm CP}$ values at odds with current global neutrino oscillation studies~\cite{deSalas:2017kay}. This led to revamped theories with viable predictions~\cite{Morisi:2013qna}. \\[-.2cm]

In this paper we re-examine the sensitivities of the DUNE experiment in probing leptonic CP violation. Rather than considering specific full-fledged neutrino theories from first principles, here we adopt a model-independent framework based on generalized CP symmetries. These are characterized by the pattern of texture zeros of the corresponding generalized CP transformation matrices~\cite{Chen:2016ica}. 
Apart from its own theoretical significance, a study based on generalized CP symmetries is important and timely, as this class of theories can naturally yield consistent values of the oscillation parameters, while retaining predictive power~\cite{Chen:2015siy,Chen:2018eou,Chen:2018lsv}. 
The paper is structured as follows. In sections~\ref{sec:CPSymm} and~\ref{sec:text-zero} we describe the generalized CP matrices and their texture zeros, while in section~\ref{sec:Expt} we briefly describe our numerical simulation. Results are presented in section~\ref{sec:ConseDUNE} and a brief summary with further discussion is given in section~\ref{sec:discussion}.

\section{Generalized CP-symmetry}
\label{sec:CPSymm}

In this section we provide a brief overview of the generalized CP method and the phenomenological consequences of CP texture zeros. 
We will mainly follow the notation of Refs.~\cite{Chen:2014wxa,Chen:2015siy,Chen:2016ica}. We start with the definition of the generalized CP symmetries. The generalized CP transformations $X_\psi$ acting on a particular fermionic field $\psi$ are defined as follows:
\begin{eqnarray}
 \psi \stackrel{CP}{\longmapsto} i X_\psi \gamma ^0 \mathcal{C} \bar{\psi} ^T. 
 \label{eq:gcp-act}
\end{eqnarray}
The associated CP transformation matrix $X_\psi$ of Eq.\eqref{eq:gcp-act} can be deemed to be a valid symmetry of the mass term provided it satisfies:
\begin{eqnarray}
X_\psi^T m_\psi X_\psi & = & m_\psi^*,\quad \textnormal{ for Majorana fields}\,, \label{eq:cpmaj}   \\
X_\psi^\dagger M^2_\psi X_\psi & = & M_\psi^{2 *},\quad \textnormal{ for Dirac fields, where } M^2_\psi \equiv m_\psi^\dagger \, m_\psi~. \label{eq:cpdir}
\end{eqnarray}
The mass matrix $m_{\psi}$ of a Majorana fermion can be diagonalized by a unitary transformation $U_{\psi}$ e.g. for a $3 \times 3$ Majorana mass matrix we have
\begin{eqnarray}
U^T_{\psi} m_{\psi} \,U_{\psi} & = & \text{diag}(m_1,m_2,m_3),\quad \textnormal{ for Majorana fields}\,,
\label{eq:diag_maj}
\end{eqnarray}
while for Dirac fermions one needs a bi-unitary transformation to diagonalize the mass matrix. However, the ``squared mass matrix'' $M^2_{\psi}$ of a Dirac fermion can also be diagonalized by a unitary transformation. For the $3 \times 3$ Dirac case we get
\begin{eqnarray}
U^\dagger_{\psi} M^2_{\psi} U_{\psi} & = & \text{diag}(m^2_1,m^2_2,m_3^2),\quad \textnormal{for Dirac fields}\,.
\label{eq:diag_dir}
\end{eqnarray}
Note that while writing down Eq.~\eqref{eq:diag_maj} and Eq.~\eqref{eq:diag_dir} we have assumed non-degenerate masses i.e. $m_1\neq m_2\neq m_3$. 
From Eqs.~\eqref{eq:cpmaj}-\eqref{eq:diag_dir}, one can show that the unitary matrix $U_{\psi}$ diagonalizing a mass matrix invariant under a given generalized CP symmetry $X_{\psi}$, satisfies the following constraint
\begin{equation}
\label{eq:XCons}U^\dagger_{\psi} X_{\psi} U^*_{\psi}\equiv P=\left\{\begin{array}{cc}
\text{diag}(\pm 1,\pm 1,\pm 1),& \textnormal{ for Majorana fields},\\[0.1in]
\text{diag}(e^{i\delta_1}, e^{i\delta_2}, e^{i\delta_3}),& \textnormal{for Dirac fields}\,,
\end{array}
\right.
\end{equation}
where $e^{i\delta_{i}}$; $i = 1,2,3$ are independent phases~\footnote{Notice that even for Majorana fermions if some of the masses are zero then the corresponding ``$\pm$'' entry should be replaced by a complex phase.}. 
It is also easy to show that $X_{\psi}$ is always a symmetric matrix~\cite{Chen:2014wxa}. 
Owing to its symmetric nature, it can be ``Takagi-decomposed'' as
\begin{eqnarray}
\label{eq:Td}X_{\psi}=\Sigma \cdot \Sigma^T\,.
\end{eqnarray}
From Eq.~\eqref{eq:XCons} and Eq.~\eqref{eq:Td} we find that
\begin{eqnarray}
P^{-\frac{1}{2}} \, U^{\dagger}_{\psi} \, \Sigma \, \equiv \,  O_3\,,
\label{eq:orth}
\end{eqnarray}
where $O_3$ is a real $3 \times 3$ orthogonal matrix. Eq.~\eqref{eq:orth} implies that the unitary matrix $U_{\psi}$ can be written in terms of the CP matrix $\Sigma$ as 
\begin{equation}
\label{ugenn}
U_{\psi} \, = \, \Sigma O_3^T \, P^{-\frac{1}{2}}\,.
\end{equation}

The orthogonal matrix $O_3$ can be parameterized in terms of three rotation angles as follows
\begin{equation}
\label{eq:O3}
O_3  = \left(\begin{array}{ccc}
1 & 0 & 0 \\
0 & \cos\theta_1   &   \sin\theta_1 \\
0 & -\sin\theta_1  &   \cos\theta_1
\end{array}\right)
\left(\begin{array}{ccc}
\cos\theta_2   &   0    &    \sin\theta_2 \\
0   &   1   &   0 \\
-\sin\theta_2   &   0   &    \cos\theta_2
\end{array}\right)
\left(\begin{array}{ccc}
\cos\theta_3     &    \sin\theta_3    &    0 \\
-\sin\theta_3    &    \cos\theta_3    & 0   \\
0    &    0     &    1
\end{array}\right)\,.
\end{equation}
where the three angles $\theta_i$; $i = 1,2,3$ are free parameters whose values can be constrained by experiments. In the rest of this work
 we will impose generalized CP symmetries on the neutrino mass matrix in order to constrain leptonic mixing parameters.
For definiteness we work in the charged lepton diagonal basis.

Before moving forward we should remark on some salient features of the generalized CP symmetry approach. 
The first thing to note is that the fermion masses are not constrained by the generalized CP symmetries and in almost all cases free parameters can be chosen to match the observed masses. The real predictive power of generalized CP symmetries lies in their ability to predict the fermion mixing angles and phases.

Another point to understand is the difference between the cases when neutrinos are Dirac or Majorana in nature. In this work we will mainly assume neutrinos to be Majorana particles, though that need not be the case \cite{Ma:2014qra,Ma:2015mjd,Chulia:2016ngi,CentellesChulia:2018gwr,CentellesChulia:2018bkz}.
If neutrinos are Dirac particles then it can be shown that \cite{Chen:2018lsv}:
\begin{itemize}
\item The mixing parameters testable in neutrino oscillations experiments, i.e., solar ($\theta_{12}$), reactor ($\theta_{13}$) and atmospheric ($\theta_{23}$) angles, as well as the CP phase ($\delta_{\rm CP}$) will be identical both for Majorana or Dirac neutrinos.
\item For Majorana neutrinos, the generalized CP symmetries in general also impose non-trivial constraints on the ``Majorana phases''. In the Dirac neutrino case, the Majorana phases are unphysical and can be rotated away by appropriate field redefinitions~\cite{Schechter:1980gr}. 
 \end{itemize}
It follows that all of our discussion in this work regarding the potential of the DUNE experiment to probe CP texture zeros holds equally well for the case of Dirac neutrinos. This is to be expected as it is well-known that oscillation experiments are insensitive to the Dirac or Majorana nature of neutrinos~\cite{Schechter:1980gk}.
 
 
\section{Texture zeros of generalized CP matrices} 
 \label{sec:text-zero}
 

In this section we look at the phenomenological consequences of all the possible texture zeros allowed in the generalized CP matrix $X$. 
As we will discuss, the presence of texture zeros in $X$ has implications for both oscillation parameters as well as for the Majorana phases, leading to implications also for neutrinoless double beta decay experiments. 
Since in this work we are mainly interested in the potential of DUNE to probe CP texture zeros, we will only focus on the implications on parameters relevant to neutrino oscillations, namely the three angles and the Dirac CP phase $\delta_{\rm CP}$. 

The $X$ matrices can be classified by the number of the textures zeros present in a given $X$. Notice that any CP matrix $X$ with more than four texture zeros is not phenomenologically viable. The various possible texture zeros matrices with four or lesser texture zeros are listed in Tab.~(\ref{tab:cp_zeros}). 
Below we briefly summarize their predictions for oscillation parameters. For further details and other 
implications of CP texture zeros see Ref.~\cite{Chen:2016ica} whose results we now summarize.

\subsubsection{Four Texture Zeros}
\label{sec:four-zero}

As mentioned before, the maximum number phenomenologically viable texture zeros $X$ can have is four. There are in total three possible four texture zero CP matrices as shown in Tab.~(\ref{tab:cp_zeros}).
However, among them only one $X_I$ is consistent with current neutrino oscillation data. This CP matrix is nothing but the CP matrix corresponding to the famous $\mu -\tau$ symmetry~\cite{Harrison:2002et,Grimus:2003yn}. It predicts:
\begin{eqnarray}
\sin^2 \theta_{13} \, = \, \sin^2\theta_2,\quad  \sin^2 \theta_{23} \, = \, \frac{1}{2}, \quad
\sin^2\theta_{12} \, = \, \sin^2\theta_3, \quad  \sin \delta_{\rm CP} \, = \, \text{sign}(\sin\theta_2 \sin2\theta_3)
 \label{x1}
\end{eqnarray}
where the parameters $\theta_i$; $i = 1,2,3$ are the angles of the orthogonal matrix $O_3$ in the parameterization of Eq.~\eqref{eq:O3}. Notice that $X_I$ predicts maximal value of atmospheric angle as well as maximal value of Dirac phase i.e. $\sin^2 \theta_{23} = 0.5$ and $\delta_{\rm CP} = \pm \pi/2 $.  

%
\begin{table}[!htbp]
\centering
\scriptsize
\addtolength{\tabcolsep}{5pt} 
 \begin{center}
  \begin{tabular}{|c|c|c| } \hline \hline
  \multicolumn{3}{|c|}{  \textbf{\begin{large} Possible CP Texture Zeros \end{large}} }\\ \hline \hline
   \multicolumn{3}{|c|}{ Four Texture Zeros } \\ \hline \hline
    ~Type~ & ${\bf X}$ & ${\bf \Sigma}$ 
    \\ \hline 
    \texttt{I}   &
     $\left( \begin{array}{ccc}
      e^{i \alpha} & 0 & 0 \\
      0 & 0 & e^{i \beta} \\
      0 & e^{i \beta} & 0
     \end{array}\right)$ &
     $\begin{pmatrix}
      e^{i\alpha/2} & 0 & 0 \\
      0 & \frac{ e^{i \beta/2} }{ \sqrt{2} } &
       i \frac{ e^{i\beta/2}  }{ \sqrt{2} } \\
      0 & \frac{ e^{i \beta/2} }{ \sqrt{2} } &
      -i\frac{ e^{i\beta/2} }{ \sqrt{2} }
     \end{pmatrix}$  
     \\ \hline
    \texttt{II} &
     $\begin{pmatrix}
      0 & 0   &  e^{i\beta}\\
      0 & e^{i\alpha}    &  0  \\
      e^{i\beta}   &  0  &  0
     \end{pmatrix}$ &
     $\begin{pmatrix}
      0 & \frac{ e^{i\beta/2} }{ \sqrt{2} } &
       i \frac{ e^{i\beta/2} }{ \sqrt{2} } \\
      e^{i\alpha/2} & 0 & 0 \\
      0 & \frac{ e^{i\beta/2} }{ \sqrt{2} } &
       -i \frac{ e^{i\beta/2} }{ \sqrt{2} }
     \end{pmatrix}$
     \\ \hline
     \texttt{III} &
     $\begin{pmatrix}
      0 & e^{i\beta} & 0 \\
      e^{i\beta} & 0 & 0 \\
      0 & 0 & e^{i\alpha}
     \end{pmatrix}$  &
     $\begin{pmatrix}
      0 & \frac{ e^{i\beta/2} }{\sqrt{2}} &
       i \frac{ e^{i\beta/2} }{\sqrt{2}} \\
      0 & \frac{ e^{i\beta/2} }{\sqrt{2}} &
       -i \frac{ e^{i\beta/2} }{\sqrt{2}} \\
      e^{i\alpha/2} & 0 & 0
     \end{pmatrix}$ 
\\ \hline \hline
   \multicolumn{3}{|c|}{ Three Texture Zeros } \\ \hline \hline
     \texttt{IV} &
     $\begin{pmatrix}
      e^{i \alpha} & 0 & 0 \\
      0 & e^{i\beta}  & 0 \\
      0 & 0 & e^{i\gamma}
     \end{pmatrix}$  &
     $\begin{pmatrix}
      e^{i \alpha/2} & 0 & 0 \\
      0 & e^{i\beta/2}  & 0 \\
      0 & 0 & e^{i\gamma/2}
     \end{pmatrix}$  
     \\ \hline \hline
   \multicolumn{3}{|c|}{Two Texture Zeros} \\ \hline \hline
     \texttt{ \uppercase \expandafter{\romannumeral5} } &
   $\begin{pmatrix}
    e^{i\alpha} & 0 & 0 \\
    0 &  e^{i\beta} c_\Theta &
    i e^{ i ( \beta + \gamma )/2 } s_\Theta \\
    0 & ie^{i ( \beta + \gamma )/2 } s_\Theta &
    e^{i\gamma} c_\Theta
   \end{pmatrix}$ &
   $\begin{pmatrix}
    e^{\frac{i \alpha}{2}} & 0 & 0 \\
    0 & e^{\frac{i\beta}{2}} c_ \frac{\Theta}{2} &
     i e^{\frac{i\beta}{2}} s_\frac{\Theta}{2} \\
    0 & i e^{\frac{i\gamma}{2}} s_\frac{\Theta}{2} &
     e^{\frac{i\gamma}{2}} c_\frac{\Theta}{2}
   \end{pmatrix}$ 
   \\ \hline
   \texttt{\uppercase\expandafter{\romannumeral6}} &
   $\begin{pmatrix}
    e^{i\alpha} c_\Theta & 0 &
     i e^{i ( \alpha + \gamma )/2 } s_\Theta \\
    0 & e^{i\beta} & 0 \\
    i e^{ ( \alpha + \gamma )/2 } s_\Theta & 0 &
     e^{i\gamma} c_ \Theta
   \end{pmatrix}$ &
   $\begin{pmatrix}
    0 & e^{ \frac{i\alpha}{2} } c_\frac{\Theta}{2} &
     i e^{ \frac{i\alpha}{2} } s_\frac{\Theta}{2} \\
    e^{ \frac{i\beta}{2} } & 0 & 0 \\
    0 & i e^{ \frac{i\gamma}{2} } s_ \frac{\Theta}{2} &
     e^{\frac{i\gamma}{2}} c_\frac{\Theta}{2}
   \end{pmatrix}$ 
    \\ \hline
   \texttt{\uppercase\expandafter{\romannumeral7}} &
   $\begin{pmatrix}
    e^{i\alpha} c_\Theta & i e^{i( \alpha + \beta )/2} s_\Theta & 0 \\
    ie^{i( \alpha + \beta )/2} s_\Theta & e^{i\beta} c_\Theta & 0 \\
    0 & 0 & e^{i\gamma}
   \end{pmatrix}$ &
   $\begin{pmatrix}
    0 & e^{\frac{i\alpha}{2}} c_\frac{\Theta}{2} &
     i e^{\frac{i\alpha}{2}} s_\frac{\Theta}{2} \\
    0 & ie^{\frac{i\beta}{2}} s_\frac{\Theta}{2} &
    e^{\frac{i\beta}{2}} c_\frac{\Theta}{2} \\
    e^{\frac{i\gamma}{2}} & 0 & 0
   \end{pmatrix}$ 
   \\ \hline \hline
   \multicolumn{3}{|c|}{One Texture Zero} \\ \hline \hline
    \texttt{\uppercase\expandafter{\romannumeral8}} &
   $\begin{pmatrix}
    0 & e^{i\alpha} c_\Theta & e^{i\beta} s_\Theta \\ \vspace{2mm}
    e^{i\alpha} c_\Theta & e^{i\gamma} s^2_\Theta &
     -e^{i\alpha_1 } c_\Theta s_\Theta  \\
    e^{i\beta} s_\Theta &
     -e^{i\alpha_1 } c_\Theta s_\Theta    &
     e^{2i{\alpha}_2 } c^2_\Theta
   \end{pmatrix}$ &
   $\frac{1}{\sqrt{2}}
    \begin{pmatrix}\vspace{2mm}
     - i e^{ i\left(\alpha - \gamma/2 \right) } &
      e^{ i\left(\alpha - \gamma/2 \right) } & 0 \\ \vspace{2mm}
     i e^{ i \gamma/2  } c_\Theta &
      e^{i\gamma/2}c_\Theta &
      \sqrt{2} e^{ i\gamma/2 } s_\Theta \\
     i e^{ i \alpha_2} s_\Theta &
      e^{ i \alpha_2} s_\Theta &
      - \sqrt{2} e^{i \alpha_2 }
      c_\Theta
   \end{pmatrix}$ \\ \hline
   \texttt{\uppercase\expandafter{\romannumeral9}} &
   $\begin{pmatrix}
    e^{i\alpha}c^2_\Theta & e^{i\beta} s_\Theta &
     e^{i\gamma} c_\Theta s_\Theta \\
    e^{i\beta} s_\Theta & 0 &
     - e^{ i\alpha_1 } c_\Theta \\
    e^{i\gamma} c_\Theta s_\Theta &
     - e^{i\alpha_1 } c_\Theta &
     e^{i( -\alpha + 2\gamma ) } s^2_\Theta
   \end{pmatrix}$ &
   $\frac{1}{\sqrt{2}}
   \begin{pmatrix}
    -i e^{i\alpha/2}s_\Theta & - e^{i\alpha/2} s_\Theta &
     \sqrt{2} e^{ i\alpha/2 } c_\Theta \\
    i e^{ i \left( \beta - \alpha/2 \right)} &
     - e^{i \left( \beta - \alpha/2 \right)} &  0 \\
    i e^{ i \left( \gamma - \alpha/2 \right)} c_\Theta &
     e^{ i \left( \gamma - \alpha/2 \right) } c_\Theta &
    \sqrt{2}e^{i\left( \gamma - \alpha/2 \right)} s_\Theta
   \end{pmatrix}$ \\ \hline
   \texttt{\uppercase\expandafter{\romannumeral10}} &
   $\begin{pmatrix}
    e^{i\alpha} c^2_\Theta & e^{i\gamma} c_\Theta s_\Theta &
     e^{i\beta} s_\Theta \\
    e^{i\gamma} c_\Theta s_\Theta &
     e^{i( -\alpha + 2\gamma) } s^2_\Theta &
     - e^{ i\alpha_1 } c_\Theta  \\
    e^{i\beta} s_\Theta &
     - e^{ i\alpha_1} c_\Theta & 0
   \end{pmatrix}$ &
   $\frac{1}{\sqrt{2}}
   \begin{pmatrix}
    -ie^{i\alpha/2} s_\Theta &
     -e^{i\alpha/2}s_\Theta &
     \sqrt{2} e^{ i\alpha/2 } c_\Theta \\
    ie^{i \left( \gamma - \alpha/2 \right) } c_\Theta &
     e^{i \left( \gamma - \alpha/2 \right)} c_\Theta &
     \sqrt{2} e^{i \left( \gamma - \alpha/2 \right)} s_\Theta \\
    ie^{i \left( \beta - \alpha/2 \right) } &
    - e^{ i \left( \beta - \alpha/2 \right) } & 0
   \end{pmatrix}$ \\ \hline\hline
   \multicolumn{3}{|c|}{Democratic CP Matrix} \\ \hline \hline   
  \texttt{\uppercase\expandafter{\romannumeral11}} &
   $  \frac{1}{\sqrt{3}}
    \begin{pmatrix} 
  e^{i \alpha } & e^{i ( \frac{ \alpha + \beta }{2} + \frac{2\pi}{3} ) } &
   e^{i ( \frac{ \alpha + \gamma}{2} + \frac{2\pi}{3} ) } \\
  e^{i ( \frac{ \alpha + \beta }{2} + \frac{2\pi}{3} ) } & e^{i\beta} &
   e^{i ( \frac{ \beta + \gamma }{2} + \frac{2\pi}{3} ) } \\
  e^{i ( \frac{ \alpha + \gamma }{2} + \frac{2\pi}{3} ) } &
   e^{i ( \frac{ \beta + \gamma }{2} + \frac{2\pi}{3} ) } & e^{i\gamma}
  \end{pmatrix}$ &
 $\text{diag}( e^{i\alpha/2}, e^{i\beta/2}, e^{i\gamma/2} ) \, e^{-\frac{i\pi}{12}}
 \begin{pmatrix}
  \sqrt{\frac{2}{3}}  & \frac{1}{\sqrt{3}} & 0 \\
  \frac{-1}{\sqrt{6}} & \frac{1}{\sqrt{3}} & \frac{1}{\sqrt{2}}  \\
  \frac{-1}{\sqrt{6}} & \frac{1}{\sqrt{3}} & \frac{-1}{\sqrt{2}} \\
  \end{pmatrix}
 \text{diag}(1, e^{i\pi/3}, 1) $
\\ \hline\hline
  \end{tabular}
 \end{center}
  \renewcommand{\arraystretch}{1.0}
 \caption{\footnotesize Possible CP transformation matrices with their corresponding ${\bf \Sigma}$ matrices. Here $\Theta$, $\alpha$, $\beta$ and $\gamma$ are real free parameters characterizing these CP transformations. We adopt short-hand notations $c_\Theta \equiv \cos \Theta$, $s_\Theta \equiv \sin \Theta, \alpha_1 \equiv - \alpha + \beta + \gamma$ and $  \alpha_2 \equiv - \alpha + \beta + \gamma/2 $. Note that not all CP texture zeros are phenomenologically viable (see text for details).}
  \label{tab:cp_zeros}
\end{table}

\subsubsection{Three Texture Zeros}
\label{sec:three-zero}

\begin{figure}[!htbp]
\begin{center}
\includegraphics[height=6cm,width=8cm]{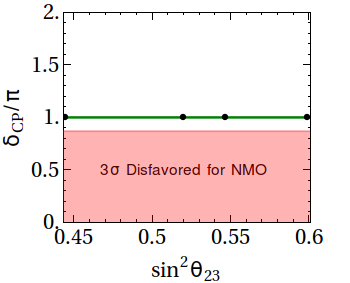}
 \end{center}
\vspace{-0.6cm}        
\caption{\footnotesize The green line corresponds to the $\sin^2 \theta_{23}-\delta_{\rm CP}$ prediction for $X_{IV}$ given in Eq.~\eqref{x4}. 
The black-points correspond to the benchmark values, of Table. \ref{tab:BenchMark} that lie within $X_{IV}$. 
The (red) shaded region is currently disfavored at 3$ \sigma $ for normal mass ordering (NMO)~\cite{deSalas:2017kay}.}
\label{fig:type4Corr}
\end{figure}

There is only one possibility for three texture zeros in the $X$ matrix, as shown in Tab.~(\ref{tab:cp_zeros}). For the oscillation parameters it predicts:
\begin{eqnarray}
\sin^2 \theta_{13} \, = \, \sin^2 \theta_{2}, \quad \sin^2 \theta_{12} \, = \, \sin^2 \theta_{3}, \quad
\sin^2 \theta_{23} \, = \, \sin^2 \theta_{1},  \quad \sin \delta_{\rm CP} \, = \,  0
\label{x4}
\end{eqnarray}
where, as before, $\theta_i$; $i = 1,2,3$ are the angles of the orthogonal matrix $O_3$ in Eq.~\eqref{eq:O3}.
Thus the three texture zero matrix $X_{IV}$ predict no CP violation. Moreover, all the mixing angles in this case remain unconstrained. Allowed parameter space of  $\sin^2 \theta_{23}-\delta_{\rm CP}$ for  this type are shown in Fig.~\ref{fig:type4Corr} by the green line. 
In plotting Fig.~\ref{fig:type4Corr} we have varied the free parameters $\theta_i$; $i =1,2,3$ in their full allowed range of ($ 0, 2\pi $],  to obtain the mixing parameters through the relations given in \eqref{x4}.
The $X_{IV}$ predict range shown in Fig.~\ref{fig:type4Corr} is then obtained by requiring that all the  mixing parameters should lie within their current 3$\sigma$ 
ranges~\cite{deSalas:2017kay}. 
The same procedure is followed throughout this section to obtain the correlation plots for $\sin^2 \theta_{23}$ and $\delta_{\rm CP}$.

\subsubsection{Two Texture Zeros}
\label{sec:two-zero}

For two texture zeros there are in total three possibilities, as listed in Tab.~(\ref{tab:cp_zeros}), all of which are phenomenologically viable. 
The first case of two texture zeros $X_V$ is nothing but generalized $\mu-\tau$ symmetry discussed in \cite{Chen:2015siy}\footnote{Generalized $\mu-\tau$ symmetry in the charged lepton sector is discussed in \cite{Chen:2018lsv}. Recently, a similar pattern using residual discrete symmetries has also been considered in \cite{Joshipura:2018cpb}.}. For oscillation parameters it predicts:
\begin{eqnarray}
\sin^2 \theta_{13} \, &=& \, \sin^2 \theta_{2}, \, \, \, \sin^2 \theta_{12} \, = \, \sin^2 \theta_{3} \;, \nonumber \\
\sin^2 \theta_{23} \,&  = & \,  \frac{1}{2} \left( 1 - \cos \Theta \cos 2\theta_1 \right), \,\,\, 
\sin\delta_{\rm CP} \, = \, \frac{ \text{sign} \left( \sin \theta_{2} \sin 2\theta_{3} \right) \sin \Theta }
{ \sqrt{ 1 - \cos^2 \Theta \cos^2 2\theta_{1} } } \;.
 \label{x5}
\end{eqnarray}

\begin{figure}[!htbp]
\begin{center}
\includegraphics[height=6cm,width=8cm]{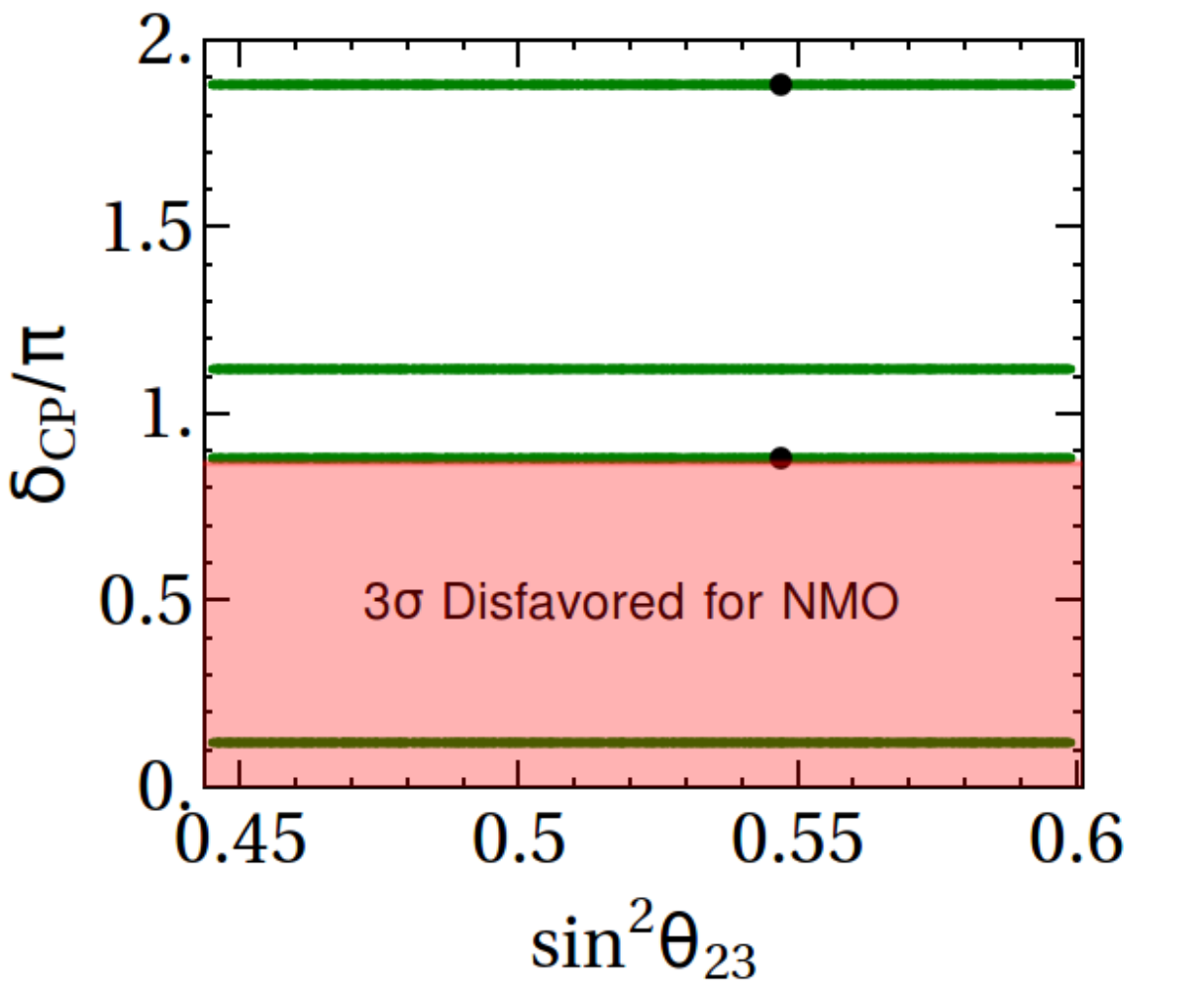}
 \end{center}
\vspace{-0.6cm}        
\caption{\footnotesize The green lines correspond to the $\sin^2 \theta_{23}-\delta_{\rm CP}$ prediction for $X_{V}$ given in Eq.~\eqref{x5}. 
The black-points correspond to the benchmark values, of Table. \ref{tab:BenchMark} that lie within $X_{V}$. 
The (red) shaded region is currently disfavored at 3$ \sigma $ for normal mass ordering (NMO)~\cite{deSalas:2017kay}.}
\label{fig:type5Corr}
\end{figure}

These expressions lead to a correlation,
\begin{equation} \label{eq:GenMutau}
\sin^2 \delta_{\rm CP} \sin^2 2\theta_{23} \, = \, \sin^2 \Theta \;,
\end{equation} 
where $\Theta$ is a label that defines the CP scenario.
 Within a given model the value of $\Theta$ is fixed. For example,
taking  $\Theta = \pi/2$ in $X_V$ leads to exact $ \mu-\tau $ reflection symmetry corresponding to the $X_I$ case.
Taking $\Theta = 0$ leads to  CP-conserving value of the Dirac CP-phase \footnote{Note that 
$ \delta_{\rm CP} = 0$ is now disfavoured  by current data at 3$ \sigma $ whereas 
$ \delta_{\rm CP} =  \pi $ is allowed at $2 \sigma $ \cite{deSalas:2017kay}.} and allows 3$ \sigma $ range of $  \theta_{23}$.
The $\sin^2 \theta_{23}-\delta_{\rm CP}$ predictions for $\Theta = 2\pi/17$ are shown in Fig.~\ref{fig:type5Corr}.
For sake of consistency and for ease of comparison, throughout this section we will take the representative value of $\Theta = 2\pi/17$ for showing the $\sin^2 \theta_{23}-\delta_{\rm CP}$ predictions. 

The second case of two texture zeros $X_{VI}$ predicts:
\begin{eqnarray}
\centering
\scriptsize
& & \sin^2 \theta_{13} =
  \frac{1}{2} \left( 1 - \cos \Theta \cos 2\theta_1 \right)
  \cos^2 \theta_2,\quad
 \sin^2 \theta_{23} =
  \frac{ 2 \sin^2 \theta_2 }{ 2 - \left( 1 - \cos \Theta \cos 2\theta_1
  \right) \cos^2 \theta_2 }\,, \nonumber \\ 
& & \sin^2 \theta_{12} = \frac{ ( 1 + \cos \Theta \cos 2\theta_1 )
  \cos^2 \theta_3 + \sin \theta_2 \left[ ( 1 - \cos \Theta \cos 2\theta_1)
  \sin \theta_2 \sin^2 \theta_3 - \cos \Theta \sin 2\theta_1 \sin 2\theta_3
  \right] }{ 2 - \left( 1 - \cos \Theta \cos 2\theta_1 \right) \cos^2 \theta_2} , \nonumber \\
& &  J_{\text{CP}} = - \frac{1}{4} \sin \Theta \sin \theta_2 \cos^2 \theta_2 \sin 2\theta_3 \, . 
 \label{x6}
\end{eqnarray}
where $J_{\text{CP}} = \frac{1}{8} \sin 2 \theta_{12} \, \sin 2 \theta_{23} \, \sin 2 \theta_{13}\, \cos\theta_{13} \,\sin \delta_{ \text{CP} }$ is the standard CP invariant.

\begin{figure}[!htbp]
\centering
\captionsetup{justification=centering,margin=2cm}
\begin{center}
\includegraphics[height=6cm,width=8cm]{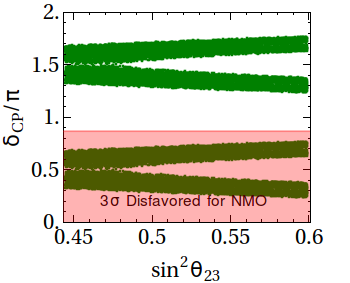}
\includegraphics[height=6cm,width=8cm]{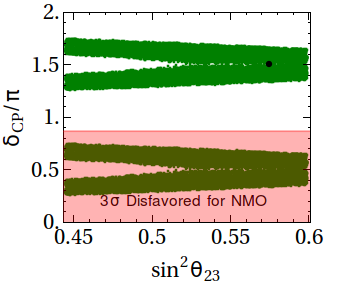}
 \end{center}
\vspace{-0.6cm}        
\caption{\footnotesize Same as Fig.~\ref{fig:type5Corr} but for $X_{VI}$ (left panel) and $X_{VII}$ (right panel).}
\label{fig:type6-7Corr}
\end{figure}

The third possibility of two texture zeros $X_{VII}$ is actually related with the $X_{VI}$ case by the exchange of the second and third rows. 
Thus the oscillation parameters in this case are same as for $X_{VI}$ case given in Eq.~\eqref{x6}, except that the angle $\theta_{23}$ and the Dirac phase $\delta_{\rm CP}$ of Eq.~\eqref{x6} should be replaced by $\pi/2 - \theta_{23}$ and $\pi + \delta_{\rm CP}$ respectively.
In Fig.~\ref{fig:type6-7Corr} we show the allowed parameter space of $\sin^2 \theta_{23}-\delta_{\rm CP}$ for $X_{VI}$ (left panel) and $X_{VII}$ (right panel) for $\Theta = 2\pi/17$.

\subsubsection{One Texture Zero}
\label{sec:one-zero}

\begin{figure}[!htbp]
\centering
\captionsetup{justification=centering,margin=2cm}
\begin{center}
\includegraphics[height=6cm,width=8cm]{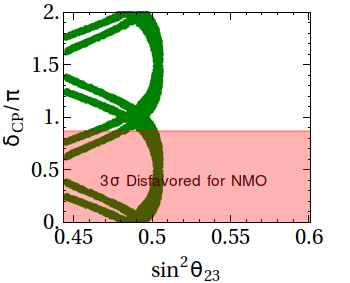}
\includegraphics[height=6cm,width=8cm]{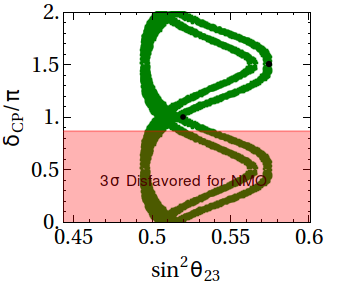}
 \end{center}
\vspace{-0.6cm}        
\caption{\footnotesize Same as Fig.~\ref{fig:type5Corr} but for $X_{IX}$ (left panel) and $X_{X}$ (right panel).}
\label{fig:type9-10-fix}
\end{figure}

For the case of one texture zero there are again three distinct possibilities, as shown in Tab.~(\ref{tab:cp_zeros}).
However, out of these three, the $X_{VIII}$ CP matrix cannot successfully account for current neutrino oscillation date. Hence, only two possibilities i.e. $X_{IX}$ and $X_X$ are phenomenologically viable.  
The CP matrix $X_{IX}$ predicts:
\begin{eqnarray}
\sin^2 \theta_{13} & = & \frac{1}{4} \left[ 1 + \cos^2 \theta_1
  \cos^2 \theta_2 + \left( 3 \cos^2 \theta_1 \cos^2 \theta_2 - 1 \right)
  \cos 2\Theta - \sqrt{2} \sin 2\theta_1 \cos^2 \theta_2 \sin 2\Theta
  \right], \nonumber \\ 
 \sin^2 \theta_{12} & = & 2 \Big[ \left( \sqrt{2} \cos \Theta \left(
  \sin \theta_1 \cos \theta_3 + \cos \theta_1 \sin \theta_2 \sin \theta_3
  \right) + \sin \Theta \left( \cos \theta_1 \cos \theta_3 - \sin \theta_1
  \sin \theta_2 \sin \theta_3 \right) \right)^2  \nonumber \\ 
 & & 
  + \cos^2 \theta_2 \sin^2 \theta_3 \sin^2 \Theta \Big] \Big/
  \Big[3 - \cos^2 \theta_1 \cos^2 \theta_2 + \left( 1 - 3
  \cos^2 \theta_1 \cos^2 \theta_2 \right) \cos 2\Theta \nonumber \\
& & 
 +  \sqrt{2}   \sin 2\theta_1 \cos^2 \theta_2 \sin 2\Theta \Big], \nonumber \\
 \sin^2\theta_{23} & = &
  \frac{ 2 \left( \sin^2 \theta_2 + \sin^2 \theta_1 \cos^2 \theta_2 \right)  }
  { 3 - \cos^2 \theta_1 \cos^2 \theta_2 + \left( 1 - 3 \cos^2 \theta_1
   \cos^2 \theta_2 \right) \cos 2\Theta + \sqrt{2} \sin 2\theta_1
   \cos^2 \theta_2 \sin 2\Theta } , \nonumber \\
J_{\text CP} & = & \frac{1}{16} \cos \theta_1 \cos \theta_2 \left[ 4
  \sin 2\theta_1 \sin \theta_2 \cos 2\theta_3 - \sin 2\theta_3 \left( 1 -
  3 \cos 2\theta_1 + 2 \cos^2 \theta_1 \cos 2\theta_2 \right) \right]
  \cos 2\Theta  \nonumber  \\
 & & 
 + \, \, \frac{1}{ 128 \sqrt{2} } \big[ \left( 12 \cos^2 \theta_1
  \sin \theta_1 \cos 3\theta_2 + \left( \sin \theta_1 - 15 \sin 3\theta_1
  \right) \cos \theta_2 \right) \sin 2\theta_3 \nonumber \\ 
  & & 
 + \, \, 4 \left( \cos\theta_1 + 3\cos 3\theta_1 \right) \sin 2\theta_2
  \cos 2\theta_3 \big] \sin 2\Theta\,.
 \label{x9}
\end{eqnarray}
 Moreover, the CP matrix $X_X$ is related to $X_{IX}$ by the exchange of the second and third rows. As a consequence, its oscillation parameters are the same as those in Eq.~\eqref{x9}, except that $\sin^2 \theta_{23} \to 1-\sin^2\theta_{23}$ and $\delta_{\rm CP} \to \pi + \delta_{\rm CP}$.
The prediction for $ \sin^2\theta_{23}$ and $\delta_{\rm CP}  $ for the CP matrices $X_X$  and  $X_{IX}$ are shown in Fig.~\ref{fig:type9-10-fix}, respectively taking $\Theta = 2\pi/17$.

\subsubsection{Democratic CP Matrix}
\label{sec:no-zero}

\begin{figure}[!htbp]
\centering
\captionsetup{justification=centering,margin=2cm}
\begin{center}
\includegraphics[height=6cm,width=8cm]{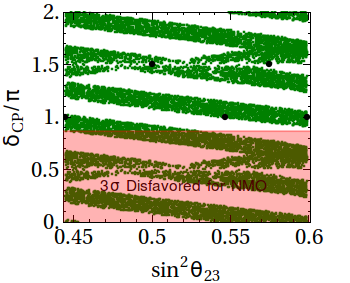}
 \end{center}
\vspace{-0.6cm}        
\caption{\footnotesize Same as Fig.~\ref{fig:type5Corr} but for $X_{XI}$.}
\label{fig:type11}
\end{figure}

Finally one can also have CP matrices without any texture zeros at all. There are infinite such possible matrices. 
A simple and predictive example of one such matrix is the so-called democratic CP matrix, as shown in Tab.~(\ref{tab:cp_zeros}). 
There are in total four such possible democratic CP matrices but it can be shown that they all lead to the same mixing parameters, up to unphysical phase redefinitions \cite{Chen:2016ica}.   
Thus, in Tab.~(\ref{tab:cp_zeros}) we have only shown one such democratic matrix, namely $X_{XI}$. It predicts:
\begin{eqnarray} \scriptsize
 \sin^2 \theta_{13} & = &
  \frac{1}{6} \left[ 4 \sin^2 \theta_2 + \sqrt{2} \sin 2\theta_2 \sin \theta_1
  + 2 \sin^2 \theta_1 \cos^2 \theta_2 \right]\,, \nonumber  \\ 
 \sin^2 \theta_{12} & = &
  \Big\{ \big[ \sin \theta_1 \big( \sqrt{2} \sin 2\theta_2 - 2 \sin \theta_1
  \sin^2 \theta_2 \big) - 4 \cos^2 \theta_2 \big] \sin^2 \theta_3
  - 2 \cos^2 \theta_1 \cos^2 \theta_3 + \big[ \sin 2\theta_1 \sin \theta_2  \nonumber   \\  
  \qquad &&
  - \sqrt{2} \cos \theta_1 \cos \theta_2 \big] \sin 2\theta_3 \Big\} /
  \big( 4 \sin^2 \theta_2 + \sqrt{2} \sin 2\theta_2 \sin \theta_1
  + 2 \sin^2 \theta_1 \cos^2 \theta_2 - 6 \big)\,, \nonumber   \\
 \sin^2 \theta_{23} & = &
  \frac{ \sin 2\theta_2 \left( \sqrt{2} \sin \theta_1 + 2 \sqrt{3}
  \cos \theta_1 \right) - 2 \sin^2 \theta_2 - \cos^2 \theta_2 \left( \sqrt{6}
  \sin 2\theta_1 + \cos 2\theta_1 + 5 \right) }{ 2 \left( 4 \sin^2 \theta_2
  + \sqrt{2} \sin 2\theta_2 \sin \theta_1 + 2 \sin^2 \theta_1 \cos^2 \theta_2
  - 6 \right)}\; , \nonumber 
\end{eqnarray}
\begin{eqnarray}  
J_{CP} & = &
  \frac{-1}{ 48 \sqrt{2} } \Big\{ \left[ 4 \sqrt{2} \sin 2\theta_2
  \sin^2 \theta_1 \cos \theta_1 + 4 \sin 2\theta_1 \cos 2\theta_2 \right]
  \cos 2\theta_3 + \Big[ 5 \sin \theta_2 \sin^2 \theta_1  \nonumber  \\
 & & \qquad
  + \sqrt{2} \sin \theta_1 \left( 5 \cos^2 \theta_1 - 1 \right) \cos \theta_2
  + \left( 3 \cos^2 \theta_1 + 1 \right) \sin 3 \theta_2 \nonumber  \\
 & & \qquad
   + \sqrt{2}
  \sin^3 \theta_1 \cos 3\theta_2 \Big] \sin 2\theta_3 \Big\}\,.  
 \label{x11}
\end{eqnarray}
In Fig.~\ref{fig:type11}, we show the prediction for $ \sin^2\theta_{23}$ and $\delta_{\rm CP}  $ corresponding to the CP matrix $X_{XI}$ with $\Theta = 2\pi/17$.

In the following sections we will look at the potential of the upcoming Deep Underground Neutrino Experiment (DUNE) to probe the predictions of the above texture zero CP matrices, focusing on the atmospheric angle $\theta_{23}$ and the Dirac CP phase $\delta_{\rm CP}$.

\section{Experimental and Simulation Details}
\label{sec:Expt}

DUNE is a proposed next generation superbeam neutrino oscillation experiment at Fermilab, USA \cite{Acciarri:2015uup, Alion:2016uaj}. The DUNE collaboration has planned to utilize existing Neutrinos from the Main Injector (NuMI) beamline design at Fermilab as a neutrino source. The far detector of DUNE will be installed at Sanford Underground Research Facility (SURF) in Lead, South Dakota. The distance from the source to the far detector is about 1300 km (810 mi) and it will be kept about 1.5 km under the surface. The first detector will record particle interactions near the source of the beam, at Fermilab, while the second, much larger, underground detector will use four $ 10 $ kton volume of liquid argon time-projection chambers (LArTPC). The expected design flux corresponds to 1.07 MW beam power which gives $1.47\times 10^{21} $ protons on target (POT) per year for an 80 GeV proton beam energy. 

For the numerical simulation of the DUNE data, we use the \texttt{GLoBES} package \cite{Huber:2004ka, Huber:2007ji} along with the required auxiliary files presented in Ref.~\cite{Alion:2016uaj}. We perform our numerical analysis considering 3.5 years running time in both neutrino and antineutrino modes.  
Throughout this work, we consider 40 kton detector volume. While performing the numerical simulation, we have also taken into account both the appearance and disappearance channels of neutrinos as well as antineutrinos.
In addition, we adopt signal and background normalization uncertainties for the appearance and disappearance channels as mentioned in the DUNE CDR~\cite{Alion:2016uaj}. 

Given that the latest global-fit values of neutrino oscillation parameters prefer normal neutrino mass ordering (i.e., $m_1 < m_2 < m_3$) over inverted neutrino mass ordering (i.e., $ m_3 < m_1 \sim m_2 $) at more than 3$ \sigma $ \cite{deSalas:2017kay, deSalas:2018bym}, we focus our study on the first scenario.
We adopt given ``theory motivated'' benchmark values for $\theta_{23}, \delta_{\rm CP} $ as true values, and determine the resulting DUNE sensitivity regions.
Concerning the former, we take the values suggested by our generalized CP theories, as characterized by the corresponding CP texture zeros.
We fix solar oscillation parameters $\theta_{12},  \Delta m^2_{21}$ at their current best-fit values in both true and test. 
On the other hand, we assume current best-fit values for $\theta_{13}$ and $ \Delta m^2_{31}$ as their true values, whereas in the test, we marginalize over their 3$ \sigma $ ranges allowed by the latest global-fit results of~\cite{deSalas:2017kay}.
In next section, we provide a detailed description of our numerical analysis for different true benchmark points in the context of DUNE.\\

\begin{figure}[!htbp]
\begin{center}
\includegraphics[height=6cm,width=6.8cm]{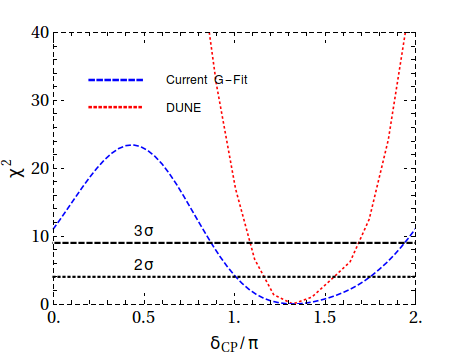}
 \end{center}
\vspace{-0.6cm}        
\caption{\footnotesize Sensitivity of DUNE vs current global-fit to probe leptonic CP violation. In order to compare the two we have taken the current current best-fit value,
$ (\sin^2\theta_{23}, \delta_{\rm CP}) = (0.547,1.32\pi) $ from \cite{deSalas:2017kay} as the true value for our simulations. The DUNE sensitivity is given after 3.5 + 3.5 years of runtime.}
\label{fig:dune-cp}
\end{figure}

Before, moving on to analyze the potential of the DUNE experiment to probe various specific 
CP texture-zero patterns, we briefly quantify DUNE's general capability in probing leptonic CP 
violation. To do that we assume, for definiteness, that the current indications from neutrino 
oscillation data survive, i.e. we fix as a benchmark the current value 
obtained in~\cite{deSalas:2017kay}.
In Fig. \ref{fig:dune-cp} we compare our current determination of the leptonic CP violating phase $\delta_{\rm CP} $ with the expected range of  
$\delta_{\rm CP} $  by the end of the 3.5 + 3.5 run period of the DUNE experiment.

From Fig. \ref{fig:dune-cp} we see that if the current best-fit CP value 
$\delta_{\rm CP} = 1.32\pi$ is indeed the true value, then after 3.5 + 3.5 years of running
the DUNE experiment there will be a significant improvement in the measurement of  
$\delta_{\rm CP}$. 
Currently  at 3$\sigma$ the $\delta_{\rm CP}$ lies in the range $(0.87 - 1.94) \pi$~\cite{deSalas:2017kay}, as indicated by the blue curve. 
However, at end of the 3.5 + 3.5 years of DUNE run the 3$\sigma$ will shrink appreciably down to $(1.08 - 1.68) \pi$, as seen in the red curve. 
Clearly the DUNE experiment offers good prospects for testing leptonic CP violation. 
Similar improvement is also expected in the measurement of the atmospheric angle $\sin^2 \theta_{23}$.
In Section \ref{sec:ConseDUNE} we investigate the capabilities of the DUNE experiment to test various 
theory predictions for $\sin^2\theta_{23}$ and $\delta_{\rm CP}$ obtained within the framework of 
generalized CP symmetry theories.

\section{Consequences of generalized CP-symmetry at DUNE}
\label{sec:ConseDUNE}

Having discussed the theory predictions originating from texture zeros in the generalized CP matrices discussed in Section \ref{sec:text-zero}, we now investigate DUNE's capability to test these predictions. 
We note that in most cases texture zeros yield regions in the $\theta_{23} - \delta_{\rm CP}$ plane, rather than a unique point~\footnote{An exception is our
texture zero CP matrix $X_I$, which corresponds to $\mu-\tau$ symmetry and predicts a unique point.}. 
Although the nature of the predicted $\theta_{23}$--$\delta_{\rm CP}$ correlations differ in each case, they often lead to overlapping regions in the $\theta_{23} - \delta_{\rm CP}$ plane. 
Hence, in this section we exploit this fact by adequately choosing benchmarks that are shared by several CP texture zero patterns. 
By employing this strategy we aim to cover all viable cases with as few representative benchmark points as possible, as listed in Tab. \ref{tab:BenchMark}. 
 These benchmark values are carefully chosen in such a way that 
\begin{enumerate}
 \item They are well-motivated from theoretical point of view and  are indeed realized in at least one of the cases discussed in Section \ref{sec:text-zero}.
\item They can be used as simulated true values against which other CP texture zero cases (those not including it) can be tested. 
\item They are such that all the CP texture zero cases of Section \ref{sec:text-zero} can be covered with as few benchmark points as possible.
\end{enumerate}

\begin{figure}[!htbp]
\begin{center}
 \begin{tabular}{lr}
\hspace{-1.5cm}
\\ \begin{subfigure}[b]{0.45\textwidth}
\includegraphics[height=6cm,width=6.8cm]{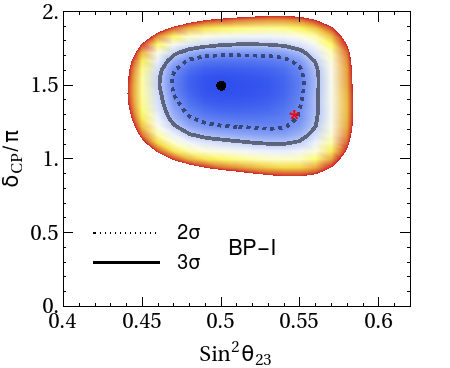}
\caption[]{BP-I: $\sin^2\theta_{23} = 0.5, \delta_{\rm CP} = 1.5\pi$.}
         \label{fig:bp1}
\end{subfigure}
~ 
\begin{subfigure}[b]{0.45\textwidth}
\includegraphics[height=6cm,width=8.0cm]{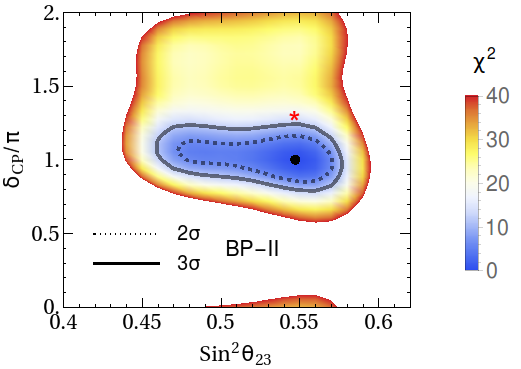}
\caption[a]{BP-II: $\sin^2\theta_{23} = 0.547, \delta_{\rm CP} = \pi$.}
         \label{fig:bp2}
\end{subfigure} \\ \\
\begin{subfigure}[b]{0.45\textwidth}
\includegraphics[height=6cm,width=6.8cm]{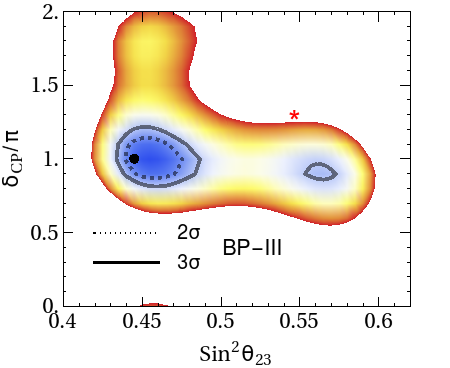}
\caption[]{BP-III: $\sin^2\theta_{23} = 0.445, \delta_{\rm CP} = \pi$.}
         \label{fig:bp3}
\end{subfigure}
~ 
\begin{subfigure}[b]{0.45\textwidth}
\includegraphics[height=6cm,width=8.0cm]{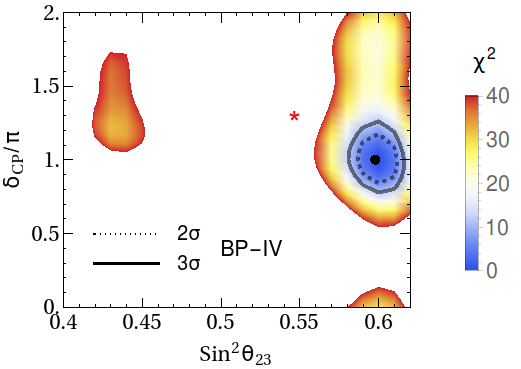}
\caption[a]{BP-IV: $\sin^2\theta_{23} = 0.599, \delta_{\rm CP} = \pi$ .}
         \label{fig:bp4}
\end{subfigure}\\ \\
\begin{subfigure}[b]{0.45\textwidth}
\includegraphics[height=6cm,width=6.8cm]{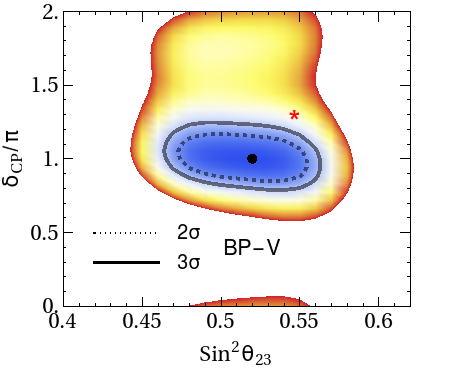}
\caption[]{BP-V: $\sin^2\theta_{23} = 0.52, \delta_{\rm CP} = \pi$.}
         \label{fig:bp5}
\end{subfigure}
~ 
\begin{subfigure}[b]{0.45\textwidth}
\includegraphics[height=6cm,width=8.0cm]{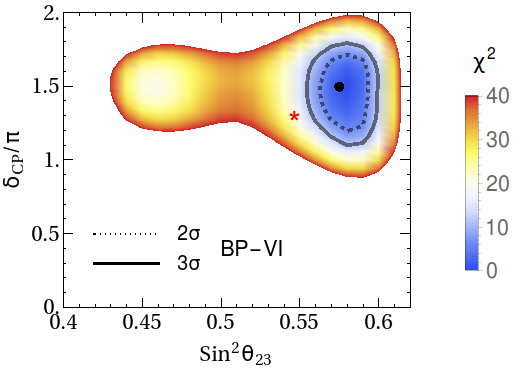}
\caption[a]{BP-VI: $\sin^2\theta_{23} = 0.575, \delta_{\rm CP} = 1.5\pi$ .}
         \label{fig:bp6}
\end{subfigure}
\end{tabular}
 \end{center}
\vspace{-0.6cm}        
\caption{\footnotesize Sensitivity regions for DUNE in the $ (\sin^2\theta_{23},\delta_{\rm CP}) $ plane for different true benchmark points (i.e., BP-I, BP-II, BP-III, BP-IV,  BP-V, and BP-VI) as given in Tab.(\ref{tab:BenchMark}). The black-dot marks represent true benchmark values of $ (\sin^2\theta_{23}, \delta_{\rm CP})$, whereas the `red-star' is the current best-fit value, $ (\sin^2\theta_{23}, \delta_{\rm CP}) = (0.547,1.32\pi) $ from \cite{deSalas:2017kay}.}
\label{fig:BMPoint}
\end{figure}

\begin{figure}[!htbp]
\begin{center}
 \begin{tabular}{lr}
\hspace{-1.5cm}
\\ \begin{subfigure}[b]{0.45\textwidth}
\includegraphics[height=6cm,width=6.8cm]{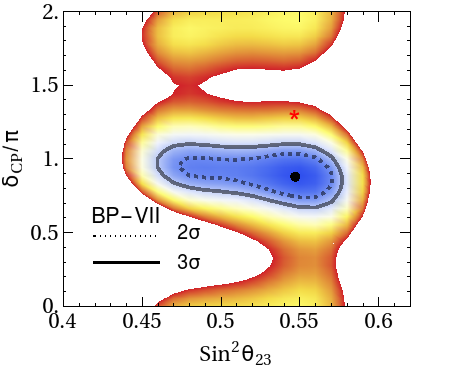}
\caption[]{BP-VII: $\sin^2\theta_{23} = 0.547, \delta_{\rm CP} = 0.88\pi$. }
         \label{fig:bp7}
\end{subfigure}
%
%
\begin{subfigure}[b]{0.45\textwidth}
\includegraphics[height=6cm,width=8.0cm]{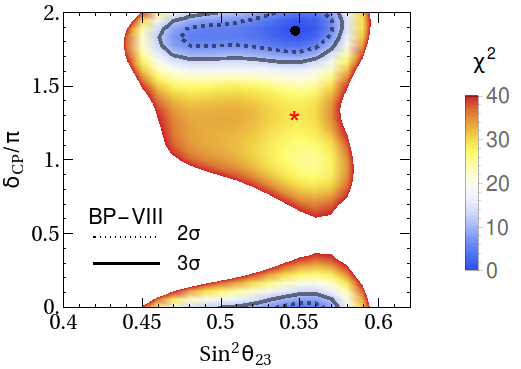}
\caption[a]{BP-VIII: $\sin^2\theta_{23} = 0.547, \delta_{\rm CP} = 1.88\pi$.}
         \label{fig:bp8}
\end{subfigure} \\ 
 \end{tabular}
 \end{center}
\vspace{-0.6cm}        
\caption{ \footnotesize DUNE sensitivity regions in the $ (\sin^2\theta_{23},\delta_{\rm CP}) $ plane for benchmark points BP-VII and BP-VIII, as given in Tab.(\ref{tab:BenchMark}). The black-dot (red-star) marks represent true benchmark (global best-fit) values of $ (\sin^2\theta_{23}, \delta_{\rm CP})$. }
\label{fig:BMPoint2}
\end{figure}

\begin{figure}[!htbp]
\begin{center}
 \begin{tabular}{lr}
\hspace{-1.5cm}
\\ \begin{subfigure}[b]{0.45\textwidth}
\includegraphics[height=6cm,width=6.8cm]{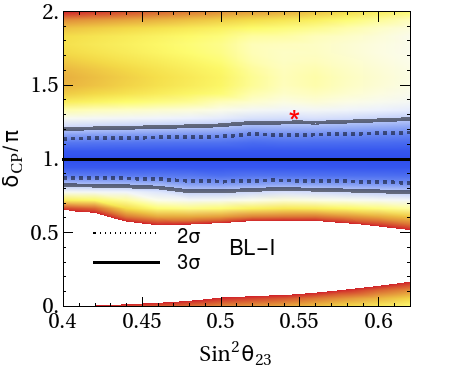}
\caption[]{BL-I: $\sin^2\theta_{23} = 0.445\oplus0.599, \delta_{\rm CP} = \pi$.}
         \label{fig:bl1}
\end{subfigure}
\begin{subfigure}[b]{0.45\textwidth}
\includegraphics[height=6cm,width=8.0cm]{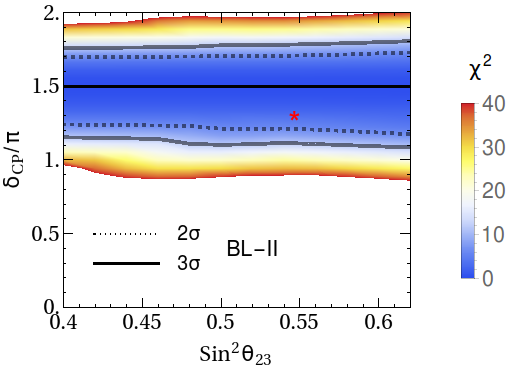}
\caption[a]{BL-II: $\sin^2\theta_{23} = 0.445\oplus0.599, \delta_{\rm CP} = 1.5\pi$.}
         \label{fig:bl2}
\end{subfigure} \\
 \end{tabular}
 \end{center}
\vspace{-0.6cm}        
\caption{\footnotesize Allowed parameter space of DUNE in $ (\sin^2\theta_{23}, \delta_{\rm CP}) $ plane for the different true benchmark line as given in Tab.(\ref{tab:BenchMark}). Also, the red-star mark represents global best-fit value of $ (\sin^2\theta_{23}, \delta_{\rm CP})$. Here, left (right) panel corresponds to CP conserving (maximal CP violating)  value of true $ \delta_{\rm CP} $. }
\label{fig:BMLine}
\end{figure}

Tab.~\ref{tab:BenchMark} summarizes our benchmark points. The second column represents the benchmark values we have adopted, while the third column indicates the generalized CP texture zero patterns in which they appear. Notice also that, while theory-motivated, all of the benchmark points in Tab.~\ref{tab:BenchMark} are also taken to lie within 3$\sigma$ of the currently allowed values~\cite{deSalas:2017kay}.
The corresponding results of our DUNE simulations are shown in Figs. \ref{fig:BMPoint}-\ref{fig:BMLine}.


 \begin{table}[htb]
        \centering 
        \addtolength{\tabcolsep}{10pt}
       \begin{tabular}{|c|c|c| }
       \hline  Scenarios    &  $(\sin^2\theta_{23}, \delta_{\rm CP})$ &  CP textures    \\ \hline \hline
       \multicolumn{3}{|c|}{ Benchmark Points }\\ \hline
       \hline \hyperref[fig:bp1]{BP-I}  &  $(0.5, 1.5\pi)$  & $X_{I},  X_{V}, X_{VI}, X_{XI} $  \\ 
       \hline \hyperref[fig:bp2]{BP-II}  & $(0.547, \pi )  $ &  $ X_{IV}, X_{V}, X_{XI} $ \\ 
       \hline \hyperref[fig:bp3]{BP-III} & $ (0.445, \pi ) $  &  $ X_{IV}, X_{V}, X_{XI} $  \\ 
       \hline \hyperref[fig:bp4]{BP-IV} & $  (0.599, \pi )$ &   $ X_{IV}, X_{V}, X_{XI} $    \\
       \hline \hyperref[fig:bp5]{BP-V} & $(0.52, \pi)$  & $X_{IV}, X_{V}, X_{X}$  \\ 
       \hline \hyperref[fig:bp6]{BP-VI} & $(0.575, 1.5\pi)$ &   $ X_{X}, X_{XI} $    \\
       \hline \hyperref[fig:bp7]{BP-VII}  & $(0.547, 0.88\pi )  $ & $ X_{V} $ \\
       \hline \hyperref[fig:bp8]{BP-VIII}  & $(0.547, 1.88\pi )  $ &  $X_{V} $ \\ \hline \hline
       \multicolumn{3}{|c|}{ Benchmark Lines }\\ \hline
       \hline \hyperref[fig:bl1]{BL-I} & $(0.445\oplus0.599, \pi)$  & $ X_{IV}, X_{V}, X_{X}, X_{XI}$  \\ 
       \hline \hyperref[fig:bl2]{BL-II} & $(0.445\oplus0.599, 1.5\pi)$ &   $ X_I, X_{V}, X_{VI}, X_{X}, X_{XI}$    \\
       \hline       
     \end{tabular}
     \caption{\footnotesize  Set of benchmark values corresponding to all the possible CP textures zeros. Note that the case BP-I represents exact $ \mu - \tau$ reflection symmetry.  Here,  abbreviation BP stands for ``benchmark point", whereas BL implies ``benchmark line".} 
     \label{tab:BenchMark}       
      \end{table} 

We now proceed to discuss the phenomenological implications of the different benchmark values as given by Tab.~\ref{tab:BenchMark}. In Fig.~\ref{fig:BMPoint} and Fig.~\ref{fig:BMPoint2}, we describe the predicted $ (\sin^2\theta_{23}, \delta_{\rm CP}) $ sensitivity regions for different benchmark points, for the case of normal neutrino mass ordering. Different color variations show $ \chi^{2} $ values ranging from $ \chi^{2} = 0$ all the way up to $ \chi^{2} = 40$ (see figure label for details). 
We have also drawn the contours for $2\sigma$ (black-dotted) and $3\sigma$ (black-solid), corresponding to $ \chi^{2} = 6.18, 11.83$ for 2 $d.o.f$, respectively.
The `black-dot' points depict the true benchmark values for different scenarios, as given by Tab.~\ref{tab:BenchMark}.
For quick comparison with our current knowledge of these parameters, in all figures, we have also shown as a red-star the current best-fit value, $ (\sin^2\theta_{23}, \delta_{\rm CP}) = (0.547,1.32\pi) $ from \cite{deSalas:2017kay}. 

Fig.~\ref{fig:bp1} shows the DUNE sensitivity regions for the BP-I scenario. Recall that this scenario corresponds to having exact $ \mu - \tau$ reflection symmetry,
and predicts the parameters $ \theta_{23}$ and $\delta_{\rm CP} $ both to be maximal. In plotting Fig.~\ref{fig:bp1} we have taken these values as true benchmark points.
One sees that after 3.5 years running time in both neutrino and antineutrino modes, DUNE will probe a large part of the $ \theta_{23}$, $\delta_{\rm CP} $ plane at the $ 3\sigma $ level. 
As shown in Tab.~\ref{tab:BenchMark}, this benchmark value is also shared by the scenarios associated to the CP matrices of $X_V, X_{VI}, X_{XI}$.
As clear from Fig.~\ref{fig:bp1}, for this BP-I case, apart from a small region, DUNE will have the capability to significantly rule out the CP conservation hypothesis, i.e $\delta_{\rm CP} = 0, \pi$.

The DUNE simulation performed by taking the second benchmark BP-II as true value is shown in Fig.~\ref{fig:bp2}. 
This benchmark point lies in the allowed ranges for the $X_{IV}, X_V, X_{XI}$ CP matrices, as shown in Tab.~\ref{tab:BenchMark}.
Since this benchmark value corresponds to CP conservation, it can test for the maximal CP violation hypothesis. As shown in Fig.~\ref{fig:bp2},  
in this case maximal CP violation can indeed be excluded at more than $3\sigma$. 
Furthermore the allowed range for the atmospheric angle can also be significantly constrained. 

Our determinations of DUNE's reach for benchmarks BP-III and BP-IV are shown in Fig.~\ref{fig:bp3} and Fig.~\ref{fig:bp4}, respectively. 
As stated in Tab.~\ref{tab:BenchMark}, these benchmarks are also shared with $X_{IV}, X_V, X_{XI}$ CP zero texture patterns. 
While both correspond to CP conserving scenarios, these benchmarks have significantly different values of the atmospheric angle, BP-III corresponding to $\sin^2 \theta_{23}$ in the lower octant ($\sin^2\theta_{23} < 0.5$), while BP-IV has $\sin^2 \theta_{23}$ in the upper
 octant ($\sin^2\theta_{23} > 0.5$).  
Just like BP-II, these benchmarks are again CP conserving, and will probe the hypothesis of maximal CP violation, which will be ruled out at more than $3\sigma$.
One also sees from these panels that DUNE will restrict the allowed $\sin^2\theta_{23}$ range to a small region around its true value. 
As shown in Fig.~\ref{fig:bp3}, for BP-III DUNE will exclude maximal atmospheric mixing to a very high significance, while the whole higher 
octant, will be disfavored at $ 2\sigma $. A small region around $ \sin^2\theta_{23} \sim 0.56 $ will remain allowed at 3$\sigma$.
For BP-IV one sees, from Fig.~\ref{fig:bp4}, that DUNE will rule out maximal atmospheric angle at $ \chi^{2} \sim 40$. Moreover, 
DUNE will also exclude the whole lower octant, $ \theta_{23} $ at more than $ 3\sigma $. 
For such benchmark values, the exact $ \mu - \tau$ reflection symmetry, i.e.  $ (\sin^2\theta_{23}, \delta_{\rm CP}) = (0.5,1.5\pi) $ will
 be excluded by DUNE at $ \chi^{2} \sim 40$.

The results of our DUNE simulations for benchmarks BP-V and BP-VI are shown in Fig.~\ref{fig:bp5} and Fig.~\ref{fig:bp6}, respectively. 
As shown in Tab.~\ref{tab:BenchMark} these benchmark points are shared by the CP matrices $ X_X, X_{XI}$.
To quantify DUNE's sensitivities in this case,  in Fig.~\ref{fig:bp5} and Fig.~\ref{fig:bp6} we adopt CP conserving and maximal CP violating true benchmark values, respectively.
Note that benchmark BP-V is CP conserving, with $\sin^2 \theta_{23} = 0.52$. Maximal CP violation can be ruled out to a very high significance in this case, 
whereas the maximal atmospheric mixing is allowed within $2 \sigma$, see Fig.~\ref{fig:bp5}. 
In contrast, benchmark BP-VI has maximal CP violation ($\delta_{\rm CP} = 1.5 \pi$) and non-maximal atmospheric mixing ($\sin^2 \theta_{23} = 0.575$).
As seen from Fig.~\ref{fig:bp6} in this case DUNE will have the potential to exclude the CP conservation hypothesis at $ \chi^{2} \sim 40$.
We also notice from BP-VI (see Fig.~\ref{fig:bp6}) that DUNE will be able to rule out the lower octant of the atmospheric angle  with more than $3\sigma $ sensitivity. Moreover, one sees that in this case the determination of $\theta_{23}, \delta_{\rm CP} $ improves considerably.

 
We now turn to the case of benchmarks BP-VII and BP-VIII, which appear associated to the $X_V, X_{XI}$ CP matrix patterns.
Notice that so far all the benchmark points we have considered (see Tab.~\ref{tab:BenchMark} and Fig.~\ref{fig:BMPoint}) were either CP conserving
 or violated CP maximally.
The current benchmark points BP-VII and BP-VIII violate CP nonmaximally.
 BP-VII has nonmaximal CP violation and nonmaximal atmospheric mixing, with $\delta_{\rm CP} = 0.88 \pi$ and $\sin^2 \theta_{23} = 0.547$. 
The results of our DUNE simulation for this benchmark are shown in Fig.~\ref{fig:bp7}.
In this case maximal CP violation will be excluded well above $3\sigma$. In contrast, CP conservation cannot be excluded even at 2$\sigma$. 
Likewise, maximal atmospheric mixing also cannot be ruled out. 
The results of our DUNE simulation for the final benchmark point, BP-VIII, corresponding to $\sin^2 \theta_{23} = 0.547$ and $\delta_{\rm CP} = 1.88 \pi$, are shown in Fig.~\ref{fig:bp8}. Here too we find that maximal CP violation can be ruled out at more than $3 \sigma$.
However, no CP violation as well as maximal atmospheric mixing remain allowed within $2 \sigma$.
Notice that for BP-VII (see Fig.~\ref{fig:bp7}) the exact $ \mu - \tau$ reflection symmetry can be ruled out at $ \chi^{2} \sim 40$, while for BP-VIII it can only be ruled out at the $3\sigma$ level, as seen from Fig.~\ref{fig:bp8}. Notice also that, except for Fig.~\ref{fig:bp1} which corresponds to BP-I case, the current best-fit value shown in the plots by the ``red star" lies outside the $3 \sigma$ range in all cases of Fig.~\ref{fig:BMPoint} and Fig.~\ref{fig:BMPoint2}.

 
Having discussed at length the results of our DUNE simulations for the theory-motivated benchmark points in Tab.~\ref{tab:BenchMark}, we now discuss two examples where the benchmarks are given by lines 
 \footnote{The benchmark lines are chosen because of their importance and frequent occurrence in the general literature as well as because points on these lines frequently fall within the allowed parameter space of various CP texture zero cases. }
as seen in the last two rows in Tab.~\ref{tab:BenchMark}.
The results are shown in Fig.~\ref{fig:BMLine}.
To perform the numerical analysis for Fig.~\ref{fig:BMLine}, we have taken several true-value points all along the ``true-value line''. The true-value points are taken close enough to each other so that they effectively form a true-value line. 
Then, taking one true-value point at a time, we perform our DUNE simulations for the test-values for that point. This is done by marginalizing over the test-values of  
$\theta_{23}$ in its allowed 3$ \sigma $ range and calculating minimum values of $ \chi^{2} $ (i.e. $ \chi^{2}_{min} $). Finally, we collect all the $ \chi^{2}_{min} $ for each  true-value of  $\theta_{23}$ and show the results in the ($\sin^2 \theta_{23}, \delta_{\rm CP}$)  plane.
In Fig.~\ref{fig:bl1} we have taken the CP conserving true benchmark value $ \delta_{\rm CP} = \pi$, whereas in Fig.~\ref{fig:bl2} 
we focused on the benchmark value $  \delta_{\rm CP} = 1.5\pi$ with maximal CP violation, as true value.
In both cases, 3$ \sigma $ ranges of $\theta_{23}  $ from the latest neutrino oscillation global-fit analysis have been adopted.
It can be seen from Fig.~\ref{fig:bl1} that maximal CP violation and, as a result, $\mu - \tau$ reflection symmetry, would be ruled out well above 3$\sigma $.

For the other benchmark line of Fig.\ref{fig:bl2} corresponding to maximal CP violation, we find that DUNE has the capability to rule out the possibility 
of CP conservation at $ 3\sigma $.  
Finally, we also note that the current best-fit point from the global analysis of neutrino oscillation data (marked by the red-star) is disfavored at more than $3\sigma$ for the CP conserving scenario depicted in Fig.~\ref{fig:bl1}. However, for the maximal CP violating case of Fig.~\ref{fig:bl2}, the current best fit point lies within the $ 2\sigma $ contour.

\begin{figure}[!htbp]
\centering
\begin{center}
\includegraphics[height=6cm,width=8cm]{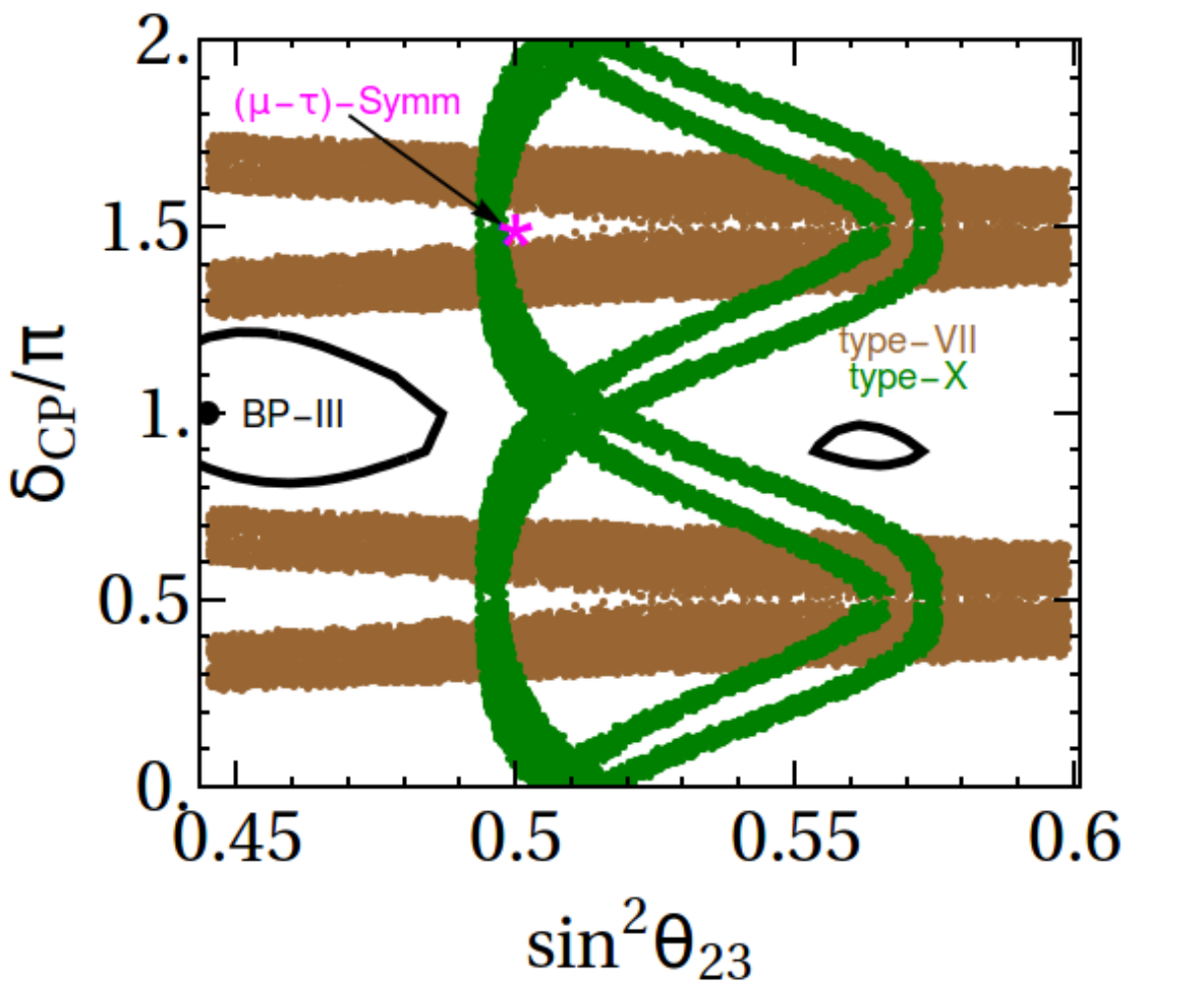}
\
\includegraphics[height=6cm,width=8cm]{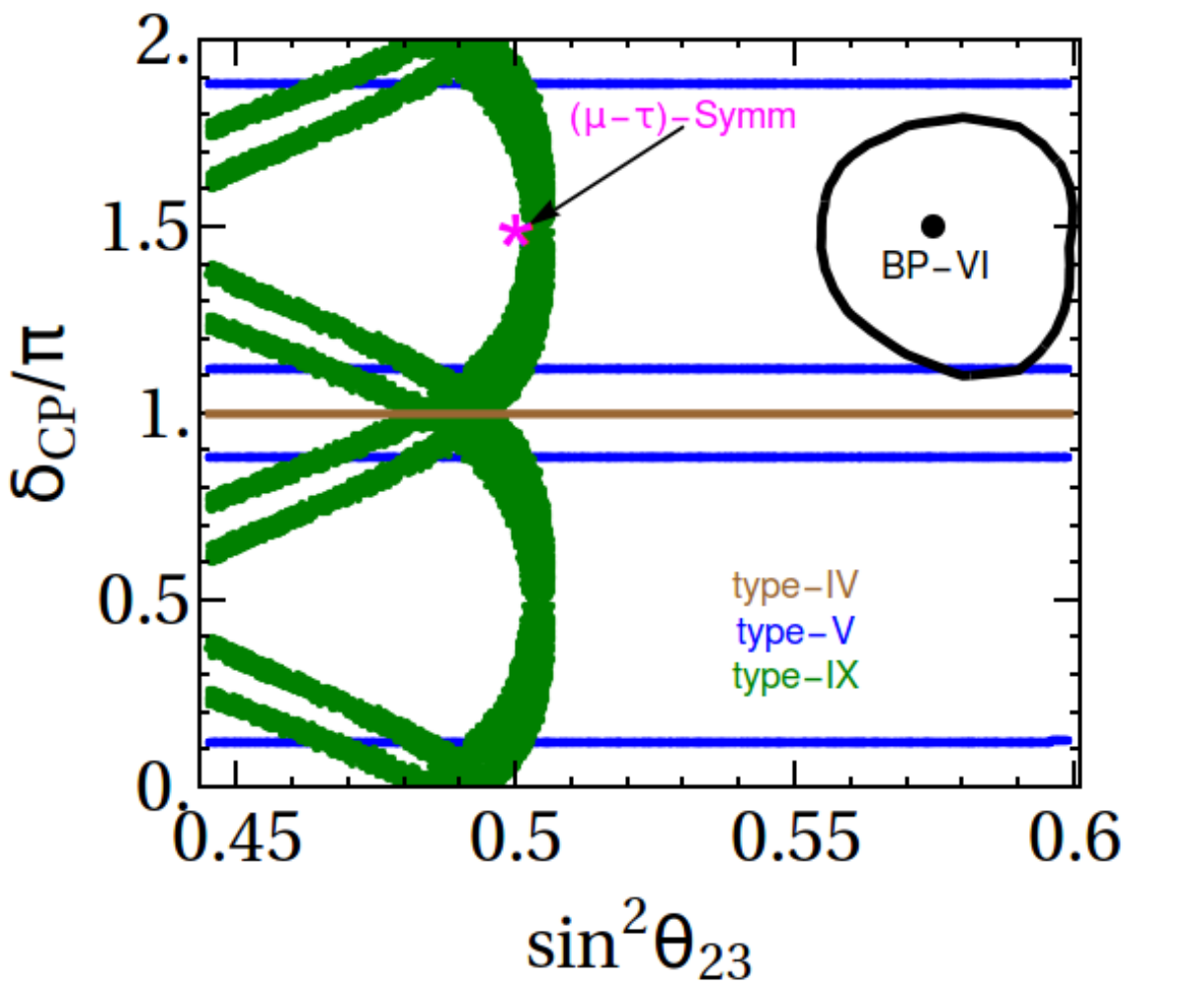}
 \end{center}
\vspace{-0.6cm}        
\caption{\footnotesize Allowed parameter space in $\sin^2\theta_{23} - \delta_{CP} $ plane  for DUNE at 3$\sigma$ for BP-III(left) and BP-VI(right) panel as shown by black contours, respectively. Also, scatter points show correlation between $\sin^2\theta_{23}  $ and $ \delta_{CP} $ for different CP matrices.}
\label{fig:TextureTest}
\end{figure}

Before concluding let us have a closer look at DUNE's discriminating power on this class of generalized CP theories. For definiteness we 
 discuss in more detail the cases of benchmark points BP-III (Fig.\ref{fig:bp3}) and BP-VI (Fig.\ref{fig:bp6}). 
This will further highlight the discriminating power of DUNE so as to probe, say, the popular mu-tau reflection symmetry ansatz.
DUNE's potential to rule out certain CP texture zero cases can be best seen by overlaying the 
plots for the allowed ranges for the CP theories on the DUNE simulated 3$\sigma$ range 
as shown in Fig. \ref{fig:TextureTest}. 
Indeed, one can see that, for any CP-motivated true value of Table. \ref{tab:BenchMark}, one can test DUNE's potential to rule out the other CP theory cases which do not include the true value.
In the left panel of Fig. \ref{fig:TextureTest} (BP-III case) one sees that, after $3.5 + 3.5$ years of run, DUNE can rule out $X_I$, $X_{VI}$, $X_{VII}$ and $X_X$ cases at more than 3$\sigma$\footnote{Note that case $X_{VI}$ is not shown in Fig.\ref{fig:TextureTest} (left panel) in order to avoid clutter. }.
In the right panel of Fig. \ref{fig:TextureTest} (BP-VI case), one sees how DUNE can rule out  $X_I$, $X_{IV}$ and $X_{IX}$ cases at more than 3$\sigma$, while only a very small part of $X_V$ survives at 3$\sigma$.
Similar plots can be drawn for all benchmark points listed in Table \ref{tab:BenchMark},
taking the DUNE simulations of Fig. \ref{fig:BMPoint} and Fig. \ref{fig:BMPoint2} and comparing them with the allowed ranges for the CP theories in Section. \ref{sec:text-zero}. 


\section{Summary and Discussion}
\label{sec:discussion}

In this paper we have focused on the potential of the DUNE experiment in probing leptonic CP violation. We have adopted the model-independent framework provided by the class of theories with generalized CP symmetries. Using their characterization by the texture zeros of the corresponding generalized CP transformation matrices we have determined the experimental sensitivities for various interesting cases. In order to illustrate our results, we have focused on the two poorly known parameters, i.e., atmospheric mixing angle $ \theta_{23} $ and the Dirac CP-phase  $\delta_{\rm CP} $, for which we have taken various benchmark values as true points in our simulations. The full set of possible benchmark values is given in Tab.~(\ref{tab:BenchMark}). We have quantified the capability of the DUNE experiment for various theory-motivated cases assuming 3.5 years of DUNE running in the neutrino mode and another 3.5 years of run in the antineutrino mode.
Our results are summarized in Figs.~\ref{fig:BMPoint}, \ref{fig:BMPoint2}, and \ref{fig:BMLine} for a representative set of benchmark points and lines.
We conclude that DUNE will be able to test such theories in a meaningful way, potentially excluding some patterns of texture zeros for the CP transformations.

Before closing we also note that, once DUNE starts taking data, the determination of $\theta_{23}$ and $\delta_{\rm CP}$ will very quickly become DUNE data driven. 
Thus, if the true value of these parameters is not near the current best fit values of \cite{deSalas:2017kay}, then the contours of the allowed regions for a given C.L. will change dramatically as shown in Fig.~\ref{fig:BMPoint} and Fig.~\ref{fig:BMPoint2}.
In order to highlight the impact of DUNE on the current state of our knowledge of $\theta_{23}-\delta_{\rm CP}$, we have marked the best fit value of the latest global fit results \cite{deSalas:2017kay} by a red colored star. 
These figures also show how the allowed region will change after 3.5+3.5 years of DUNE run, if any of the theory predicted benchmark points is indeed the true value of these parameters.
For example, for the BP-I point, at the end of 3.5+3.5 year run of DUNE, the current best fit point will be almost 2$\sigma$ away from the true value, as shown in Fig.~\ref{fig:bp1}, while for BP-II it will be more than 3$\sigma$ away, as evident from Fig.~\ref{fig:bp2}. 
Similar conclusions can also be drawn for other benchmark points, as shown in Fig.~\ref{fig:BMPoint} and in Fig.~\ref{fig:BMPoint2}. In some cases, such as for BP-III and BP-IV, the current best fit point can be ruled out at more than $\chi^{2} = 40$, as shown in Fig.~\ref{fig:bp3} and Fig.~\ref{fig:bp4}. 
Thus, if one of these benchmark points is indeed the true value, DUNE will unambiguously drive the data towards it, away from the current best fit.

\section{Acknowledgements}
We thank P. Denton for his valuable comments. The research work of N. Nath was supported in part by the National Natural Science Foundation of China under grant No. 11775231. 
The research work of R. Srivastava and J.W.F Valle was supported by the Spanish grants FPA2017-85216-P (AEI/FEDER, UE), SEV-2014-0398, PROMETEO/2018/165 (Generalitat  Valenciana) and the Spanish Red Consolider MultiDark FPA2017-90566-REDC.

 \bibliographystyle{utphys}

\begin{thebibliography}{10}

\bibitem{Kajita:2016cak}
T.~Kajita, ``{Nobel Lecture: Discovery of atmospheric neutrino oscillations},''
  \href{http://dx.doi.org/10.1103/RevModPhys.88.030501}{{\em Rev. Mod. Phys.}
  {\bfseries 88} no.~3, (2016) 030501}.

\bibitem{McDonald:2016ixn}
A.~B. McDonald, ``{Nobel Lecture: The Sudbury Neutrino Observatory: Observation
  of flavor change for solar neutrinos},''
  \href{http://dx.doi.org/10.1103/RevModPhys.88.030502}{{\em Rev. Mod. Phys.}
  {\bfseries 88} no.~3, (2016) 030502}.

\bibitem{Valle:2015pba}
J.~W.~F. Valle and J.~C. Romao, {\em {Neutrinos in high energy and
  astroparticle physics}}.
\newblock John Wiley \& Sons (2015),~www.wiley.com/buy/9783527411979.

\bibitem{Kuzmin:1985mm}
V.~A. Kuzmin, V.~A. Rubakov, and M.~E. Shaposhnikov, ``{On the Anomalous
  Electroweak Baryon Number Nonconservation in the Early Universe},''
\href{http://dx.doi.org/10.1016/0370-2693(85)91028-7}{{\em Phys. Lett.}
  {\bfseries 155B} (1985) 36}.

\bibitem{Dine:2003ax}
M.~Dine and A.~Kusenko, ``{The Origin of the matter - antimatter asymmetry},''
  \href{http://dx.doi.org/10.1103/RevModPhys.76.1}{{\em Rev.Mod.Phys.}
  {\bfseries 76} (2003) 1},
  \href{http://arxiv.org/abs/hep-ph/0303065}{{\ttfamily arXiv:hep-ph/0303065
  [hep-ph]}}.

\bibitem{Fukugita:1986hr}
M.~Fukugita and T.~Yanagida, ``{Baryogenesis Without Grand Unification},''
  \href{http://dx.doi.org/10.1016/0370-2693(86)91126-3}{{\em Phys.Lett.}
  {\bfseries B174} (1986) 45}.

\bibitem{Acciarri:2016crz}
{\bfseries DUNE} Collaboration, R.~Acciarri {\em et~al.}, ``{Long-Baseline
  Neutrino Facility (LBNF) and Deep Underground Neutrino Experiment (DUNE)
  Conceptual Design Report Volume 1: The LBNF and DUNE Projects},''
  \href{http://arxiv.org/abs/1601.05471}{{\ttfamily arXiv:1601.05471
  [physics.ins-det]}}.

\bibitem{Schechter:1980gr}
J.~Schechter and J.~W.~F. Valle, ``{Neutrino Masses in SU(2) x U(1)
  Theories},'' \href{http://dx.doi.org/10.1103/PhysRevD.22.2227}{{\em Phys.
  Rev.} {\bfseries D22} (1980) 2227}.

\bibitem{Chatterjee:2017ilf}
S.~S. Chatterjee {\em et~al.}, ``{Cornering the revamped BMV model with
  neutrino oscillation data},''
  \href{http://dx.doi.org/10.1016/j.physletb.2017.09.052}{{\em Phys. Lett.}
  {\bfseries B774} (2017) 179--182},
  \href{http://arxiv.org/abs/1708.03290}{{\ttfamily arXiv:1708.03290
  [hep-ph]}}.

\bibitem{Srivastava:2017sno}
R.~Srivastava, C.~A. Ternes, M.~Tórtola, and J.~W.~F. Valle, ``{Testing a
  lepton quarticity flavor theory of neutrino oscillations with the DUNE
  experiment},'' \href{http://dx.doi.org/10.1016/j.physletb.2018.01.014}{{\em
  Phys. Lett.} {\bfseries B778} (2018) 459--463},
\href{http://arxiv.org/abs/1711.10318}{{\ttfamily arXiv:1711.10318 [hep-ph]}}.

\bibitem{Srivastava:2018ser}
R.~Srivastava, C.~A. Ternes, M.~Tortola, and J.~W.~F. Valle, ``{Zooming in on
  neutrino oscillations with DUNE},''
  \href{http://dx.doi.org/10.1103/PhysRevD.97.095025}{{\em Phys. Rev.}
  {\bfseries D97} no.~9, (2018) 095025},
\href{http://arxiv.org/abs/1803.10247}{{\ttfamily arXiv:1803.10247 [hep-ph]}}.

\bibitem{Chakraborty:2018dew}
K.~Chakraborty, K.~N. Deepthi, S.~Goswami, A.~S. Joshipura, and N.~Nath,
  ``{Partial $\mu$-$\tau$ Reflection Symmetry and Its Verification at DUNE and
  Hyper-Kamiokande},'' \href{http://dx.doi.org/10.1103/PhysRevD.98.075031}{{\em
  Phys. Rev.} {\bfseries D98} no.~7, (2018) 075031},
\href{http://arxiv.org/abs/1804.02022}{{\ttfamily arXiv:1804.02022 [hep-ph]}}.

\bibitem{Nath:2018xkz}
N.~Nath, ``{Consequences of $ \mu-\tau $ Reflection Symmetry at DUNE},''
  \href{http://dx.doi.org/10.1103/PhysRevD.98.075015}{{\em Phys. Rev.}
  {\bfseries D98} no.~7, (2018) 075015},
\href{http://arxiv.org/abs/1805.05823}{{\ttfamily arXiv:1805.05823 [hep-ph]}}.

\bibitem{Fukuyama:1997ky}
T.~Fukuyama and H.~Nishiura, ``{Mass matrix of Majorana neutrinos},''
\href{http://arxiv.org/abs/hep-ph/9702253}{{\ttfamily arXiv:hep-ph/9702253
  [hep-ph]}}.

\bibitem{Babu:2002dz}
K.~S. Babu, E.~Ma, and J.~W.~F. Valle, ``{Underlying A(4) symmetry for the
  neutrino mass matrix and the quark mixing matrix},''
  \href{http://dx.doi.org/10.1016/S0370-2693(02)03153-2}{{\em Phys. Lett.}
  {\bfseries B552} (2003) 207--213},
  \href{http://arxiv.org/abs/hep-ph/0206292}{{\ttfamily arXiv:hep-ph/0206292
  [hep-ph]}}.

\bibitem{Harrison:2002et}
P.~F. Harrison and W.~G. Scott, ``{mu - tau reflection symmetry in lepton
  mixing and neutrino oscillations},''
  \href{http://dx.doi.org/10.1016/S0370-2693(02)02772-7}{{\em Phys. Lett.}
  {\bfseries B547} (2002) 219--228},
  \href{http://arxiv.org/abs/hep-ph/0210197}{{\ttfamily arXiv:hep-ph/0210197
  [hep-ph]}}.

\bibitem{Grimus:2003yn}
W.~Grimus and L.~Lavoura, ``{A Nonstandard CP transformation leading to maximal
  atmospheric neutrino mixing},''
  \href{http://dx.doi.org/10.1016/j.physletb.2003.10.075}{{\em Phys. Lett.}
  {\bfseries B579} (2004) 113--122},
  \href{http://arxiv.org/abs/hep-ph/0305309}{{\ttfamily arXiv:hep-ph/0305309
  [hep-ph]}}.

\bibitem{Xing:2015fdg}
Z.-z. Xing and Z.-h. Zhao, ``{A review of mu-tau flavor symmetry in neutrino
  physics},'' \href{http://dx.doi.org/10.1088/0034-4885/79/7/076201}{{\em Rept.
  Prog. Phys.} {\bfseries 79} no.~7, (2016) 076201},
\href{http://arxiv.org/abs/1512.04207}{{\ttfamily arXiv:1512.04207 [hep-ph]}}.

\bibitem{deSalas:2017kay}
P.~F. de~Salas {\em et~al.}, ``{Status of neutrino oscillations 2018: 3$\sigma$
  hint for normal mass ordering and improved CP sensitivity},''
  \href{http://dx.doi.org/10.1016/j.physletb.2018.06.019}{{\em Phys. Lett.}
  {\bfseries B782} (2018) 633--640},
  \href{http://arxiv.org/abs/1708.01186}{{\ttfamily arXiv:1708.01186
  [hep-ph]}}.
\url{http://globalfit.astroparticles.es/}.

\bibitem{Morisi:2013qna}
S.~Morisi, D.~V. Forero, J.~C. Romao, and J.~W.~F. Valle, ``{Neutrino mixing
  with revamped $A_4$ flavor symmetry},''
  \href{http://dx.doi.org/10.1103/PhysRevD.88.016003}{{\em Phys. Rev.}
  {\bfseries D88} no.~1, (2013) 016003},
\href{http://arxiv.org/abs/1305.6774}{{\ttfamily arXiv:1305.6774 [hep-ph]}}.

\bibitem{Chen:2016ica}
P.~Chen, G.-J. Ding, F.~Gonzalez-Canales, and J.~W.~F. Valle, ``{Classifying CP
  transformations according to their texture zeros: theory and implications},''
  \href{http://dx.doi.org/10.1103/PhysRevD.94.033002}{{\em Phys. Rev.}
  {\bfseries D94} no.~3, (2016) 033002},
  \href{http://arxiv.org/abs/1604.03510}{{\ttfamily arXiv:1604.03510
  [hep-ph]}}.

\bibitem{Chen:2015siy}
P.~Chen {\em et~al.}, ``{Generalized $\mu-\tau$ reflection symmetry and
  leptonic CP violation},''
  \href{http://dx.doi.org/10.1016/j.physletb.2015.12.069}{{\em Phys. Lett.}
  {\bfseries B753} (2016) 644--652},
  \href{http://arxiv.org/abs/1512.01551}{{\ttfamily arXiv:1512.01551
  [hep-ph]}}.

\bibitem{Chen:2018eou}
P.~Chen, S.~Centelles~Chulia, G.-J. Ding, R.~Srivastava, and J.~W.~F. Valle,
  ``{Realistic tribimaximal neutrino mixing},''
  \href{http://dx.doi.org/10.1103/PhysRevD.98.055019}{{\em Phys. Rev.}
  {\bfseries D98} no.~5, (2018) 055019},
\href{http://arxiv.org/abs/1806.03367}{{\ttfamily arXiv:1806.03367 [hep-ph]}}.

\bibitem{Chen:2018lsv}
P.~Chen, S.~Centelles~Chulia, G.-J. Ding, R.~Srivastava, and J.~W.~F. Valle,
  ``{Neutrino Predictions from Generalized CP Symmetries of Charged Leptons},''
  \href{http://dx.doi.org/10.1007/JHEP07(2018)077}{{\em JHEP} {\bfseries 07}
  (2018) 077},
\href{http://arxiv.org/abs/1802.04275}{{\ttfamily arXiv:1802.04275 [hep-ph]}}.

\bibitem{Chen:2014wxa}
P.~Chen, C.-C. Li, and G.-J. Ding, ``{Lepton Flavor Mixing and CP Symmetry},''
  \href{http://dx.doi.org/10.1103/PhysRevD.91.033003}{{\em Phys. Rev.}
  {\bfseries D91} (2015) 033003},
  \href{http://arxiv.org/abs/1412.8352}{{\ttfamily arXiv:1412.8352 [hep-ph]}}.

\bibitem{Ma:2014qra}
E.~Ma and R.~Srivastava, ``{Dirac or inverse seesaw neutrino masses with $B-L$
  gauge symmetry and $S_3$ flavor symmetry},''
  \href{http://dx.doi.org/10.1016/j.physletb.2014.12.049}{{\em Phys. Lett.}
  {\bfseries B741} (2015) 217--222},
  \href{http://arxiv.org/abs/1411.5042}{{\ttfamily arXiv:1411.5042 [hep-ph]}}.

\bibitem{Ma:2015mjd}
E.~Ma, N.~Pollard, R.~Srivastava, and M.~Zakeri, ``{Gauge $B-L$ Model with
  Residual $Z_3$ Symmetry},''
  \href{http://dx.doi.org/10.1016/j.physletb.2015.09.010}{{\em Phys. Lett.}
  {\bfseries B750} (2015) 135--138},
  \href{http://arxiv.org/abs/1507.03943}{{\ttfamily arXiv:1507.03943
  [hep-ph]}}.

\bibitem{Chulia:2016ngi}
S.~Centelles~Chulia {\em et~al.}, ``{Dirac Neutrinos and Dark Matter Stability
  from Lepton Quarticity},''
  \href{http://dx.doi.org/10.1016/j.physletb.2017.01.070}{{\em Phys. Lett.}
  {\bfseries B767} (2017) 209--213},
  \href{http://arxiv.org/abs/1606.04543}{{\ttfamily arXiv:1606.04543
  [hep-ph]}}.

\bibitem{CentellesChulia:2018gwr}
S.~Centelles~Chulia, R.~Srivastava, and J.~W.~F. Valle, ``{Seesaw roadmap to
  neutrino mass and dark matter},''
  \href{http://dx.doi.org/10.1016/j.physletb.2018.03.046}{{\em Phys. Lett.}
  {\bfseries B781} (2018) 122--128},
\href{http://arxiv.org/abs/1802.05722}{{\ttfamily arXiv:1802.05722 [hep-ph]}}.

\bibitem{CentellesChulia:2018bkz}
S.~Centelles~Chulia, R.~Srivastava, and J.~W.~F. Valle, ``{Seesaw Dirac
  neutrino mass through dimension-six operators},''
  \href{http://dx.doi.org/10.1103/PhysRevD.98.035009}{{\em Phys. Rev.}
  {\bfseries D98} no.~3, (2018) 035009},
\href{http://arxiv.org/abs/1804.03181}{{\ttfamily arXiv:1804.03181 [hep-ph]}}.

\bibitem{Schechter:1980gk}
J.~Schechter and J.~W.~F. Valle, ``{Neutrino Oscillation Thought Experiment},''
\href{http://dx.doi.org/10.1103/PhysRevD.23.1666}{{\em Phys. Rev.} {\bfseries
  D23} (1981) 1666}.

\bibitem{Joshipura:2018cpb}
A.~S. Joshipura and K.~M. Patel, ``{Pseudo-Dirac neutrinos from flavour
  dependent CP symmetry},''
  \href{http://dx.doi.org/10.1007/JHEP07(2018)137}{{\em JHEP} {\bfseries 07}
  (2018) 137},
\href{http://arxiv.org/abs/1805.02002}{{\ttfamily arXiv:1805.02002 [hep-ph]}}.

\bibitem{Acciarri:2015uup}
{\bfseries DUNE} Collaboration, R.~Acciarri {\em et~al.}, ``{Long-Baseline
  Neutrino Facility (LBNF) and Deep Underground Neutrino Experiment (DUNE)
  Conceptual Design Report Volume 2: The Physics Program for DUNE at LBNF},''
  \href{http://arxiv.org/abs/1512.06148}{{\ttfamily arXiv:1512.06148
  [physics.ins-det]}}.

\bibitem{Alion:2016uaj}
{\bfseries DUNE} Collaboration, T.~Alion {\em et~al.}, ``{Experiment Simulation
  Configurations Used in DUNE CDR},''
\href{http://arxiv.org/abs/1606.09550}{{\ttfamily arXiv:1606.09550
  [physics.ins-det]}}.

\bibitem{Huber:2004ka}
P.~Huber, M.~Lindner, and W.~Winter, ``{Simulation of long-baseline neutrino
  oscillation experiments with GLoBES (General Long Baseline Experiment
  Simulator)},'' \href{http://dx.doi.org/10.1016/j.cpc.2005.01.003}{{\em
  Comput. Phys. Commun.} {\bfseries 167} (2005) 195},
  \href{http://arxiv.org/abs/hep-ph/0407333}{{\ttfamily arXiv:hep-ph/0407333
  [hep-ph]}}.

\bibitem{Huber:2007ji}
P.~Huber, J.~Kopp, M.~Lindner, M.~Rolinec, and W.~Winter, ``{New features in
  the simulation of neutrino oscillation experiments with GLoBES 3.0: General
  Long Baseline Experiment Simulator},''
  \href{http://dx.doi.org/10.1016/j.cpc.2007.05.004}{{\em Comput. Phys.
  Commun.} {\bfseries 177} (2007) 432--438},
  \href{http://arxiv.org/abs/hep-ph/0701187}{{\ttfamily arXiv:hep-ph/0701187
  [hep-ph]}}.

\bibitem{deSalas:2018bym}
P.~F. De~Salas, S.~Gariazzo, O.~Mena, C.~A. Ternes, and M.~Tortola, ``{Neutrino
  Mass Ordering from Oscillations and Beyond: 2018 Status and Future
  Prospects},'' \href{http://dx.doi.org/10.3389/fspas.2018.00036}{{\em Front.
  Astron. Space Sci.} {\bfseries 5} (2018) 36},
\href{http://arxiv.org/abs/1806.11051}{{\ttfamily arXiv:1806.11051 [hep-ph]}}.

\end{thebibliography}

\providecommand{\href}[2]{#2}\begingroup\raggedright\endgroup

\end{document}